\documentclass[twocolumn,nofootinbib,prd]{revtex4-1}
\usepackage{amsmath,graphicx,amssymb,slashed,amsfonts,xcolor}
\RequirePackage[T1]{fontenc}
\usepackage[utf8x]{inputenc}
\usepackage{setspace}
\usepackage{datetime}
\settimeformat{ampmtime}
\usepackage{epsfig,multirow,float}
\RequirePackage{graphicx}
\RequirePackage{mathptmx}
\usepackage{cancel,slashed}      
\RequirePackage{flushend}
\usepackage{rotating}
\usepackage{calrsfs}
\usepackage[normalem]{ulem}

\usepackage[colorlinks,citecolor=blue,urlcolor=blue,linkcolor=blue]{hyperref}

\def\wtil#1{\widetilde{#1}}
\definecolor{greena}{rgb}{0.13, 0.55, 0.13}

\begin{document}

\title{Probing the anomalous triple gauge boson couplings in {\boldmath $e^+e^-\to W^+W^-$} using     {\boldmath$W$} polarizations with polarized beams }

\author{Rafiqul Rahaman}
\email{rr13rs033@iiserkol.ac.in}
\author{Ritesh K. Singh}
\email{ritesh.singh@iiserkol.ac.in}
\affiliation{Department of Physical Sciences, Indian Institute of Science
	Education and Research Kolkata, Mohanpur, 741246, India}
\begin{abstract}
    We study the anomalous $W^+W^-V$ ($V=\gamma,Z$) couplings 
    in $e^+e^-\to W^+W^-$  using the complete set of polarization observables of
    $W$ boson with longitudinally polarized electron ($e^-$) and positron ($e^+$)  beams.
    For the effective $W^+W^-V$ couplings, we use 
    the most general Lorentz invariant form factor parametrization as well as 
    $SU(2)\times U(1)$  invariant dimension-$6$ effective operators. 
   We estimate simultaneous limits on the anomalous
    couplings  using the Markov-Chain--Monte-Carlo (MCMC) method 
    for an $e^+e^-$ collider running at center of mass energy of $\sqrt{s}=500$ GeV and the integrated
    luminosity of ${\cal L}=100$ fb$^{-1}$, $3.2$ ab$^{-1}$ and $4$ ab$^{-1}$.
    The best limits on the anomalous couplings are obtained for $e^-$ and $e^+$ polarization being 
    $(\pm 0.8,\mp 0.6)$ for both $100$ fb$^{-1}$ and $3.2$ ab$^{-1}$ of luminosity.
\end{abstract}

\maketitle

\section{Introduction}

The non-Abelian gauge symmetry $SU(2)\times U(1)$ of the Standard Model (SM)
allows the $WWV$ ($V=\gamma, Z$) couplings after the electroweak  
symmetry breaking (EWSB) by the Higgs field discovered  at the large hadron collider 
(LHC)~\cite{Chatrchyan:2012xdj}. To test the EWSB, the $WWV$ couplings have to
be measured precisely, which is still lacking. We intend to study the measurement of
these couplings using  polarization observables of the  spin-$1$ 
boson~\cite{Bourrely:1980mr,Abbiendi:2000ei,Ots:2004hk,Boudjema:2009fz,
    Aguilar-Saavedra:2015yza,Rahaman:2016pqj,Nakamura:2017ihk}.
To test the SM $WWV$ couplings, 
one has to hypothesize beyond the SM (BSM) couplings  and make sure they do not
appear at all or are severely constrained. One approach is to  consider $SU(2)\times U(1)$
invariant  higher dimension effective
operators which provide  the $WWV$ form factors after EWSB~\cite{Buchmuller:1985jz}. The effective 
Lagrangian considering the higher dimension operators can be written as 
\begin{equation}\label{eq:eft}
    {\cal L}_{eft} = {\cal L}_{SM} + \sum_i \frac{c_i^{{\cal O}(6)}}{\Lambda^2}{\cal O}_i^{(6)} 
    + \sum_i \frac{c_i^{{\cal O}(8)}}{\Lambda^4}{\cal O}_i^{(8)} + \dots ,
\end{equation}
where $c_i^{{\cal O}(6,8)}$ are the couplings of the higher dimension operators ${\cal O}_i^{(6,8)}$
and $\Lambda$ is the energy scale below which the theory is valid.
To the lowest order 
(up to dimension-$6$) the operators contributing to  $WWV$ couplings are~\cite{Hagiwara:1993ck,Degrande:2012wf} 
\begin{eqnarray}\label{eq:opertaors-dim6}
    {\cal O}_{WWW}&=&\mbox{Tr}[W_{\mu\nu}W^{\nu\rho}W_{\rho}^{\mu}] ,\nonumber\\
    {\cal O}_W&=&(D_\mu\Phi)^\dagger W^{\mu\nu}(D_\nu\Phi) ,\nonumber\\
    {\cal O}_B&=&(D_\mu\Phi)^\dagger B^{\mu\nu}(D_\nu\Phi) ,\nonumber\\
    {\cal O}_{\wtil{WWW}}&=&\mbox{Tr}[\wtil{W}_{\mu\nu}W^{\nu\rho}W_{\rho}^{\mu}] ,\nonumber\\
    {\cal O}_{\wtil W}&=&(D_\mu\Phi)^\dagger \wtil{ W}^{\mu\nu}(D_\nu\Phi) ,
\end{eqnarray}
where $\Phi$ is the Higgs doublet field and
\begin{eqnarray}
    D_\mu &=& \partial_\mu + \frac{i}{2} g \tau^I W^I_\mu + \frac{i}{2} g^\prime B_\mu ,\nonumber \\
    W_{\mu\nu} &=& \frac{i}{2} g\tau^I (\partial_\mu W^I_\nu - \partial_\nu W^I_\mu
    + g \epsilon_{IJK} W^J_\mu W^K_\nu ) ,\nonumber\\
    B_{\mu \nu} &=& \frac{i}{2} g^\prime (\partial_\mu B_\nu - \partial_\nu B_\mu) .
\end{eqnarray}
Here $g$ and $g^\prime$ are the $SU(2)$ and $U(1)$ couplings, respectively.
Among these operators, ${\cal O}_{WWW}$,  ${\cal O}_W$ and  ${\cal O}_B$ are
$CP$-even, while ${\cal O}_{\wtil{WWW}}$  and ${\cal O}_{\wtil W}$ are
$CP$-odd. 
These effective operators, 
after EWSB, also provide  $HZV$ and  $HWW$ couplings, which can be examined in various 
processes, e.g., $ZV/ZW/HV/HW$ production processes. These  processes may contain some other effective operators  as well.
We note that the $W$ pair production process also contains anomalous couplings other than the aTGC~\cite{Zhang:2016zsp,Baglio:2019uty}. However, for simplicity, we study this process only with the anomalous gauge boson couplings.

The other alternative to step beyond the  SM $WWV$ structure is to consider   the
most  general Lorentz invariant effective form factors in a model independent way.
A Lagrangian for the above parametrization is given by~\cite{Hagiwara:1986vm}
\begin{eqnarray}
    {\cal L}_{WWV} &=&ig_{WWV}\left(g_1^V(W_{\mu\nu}^+W^{-\mu}-
    W^{+\mu}W_{\mu\nu}^-)V^\nu\right.\nonumber\\
    &+&ig_4^VW_\mu^+W^-_\nu(\partial^\mu V^\nu+\partial^\nu V^\mu)\nonumber\\
    &-&ig_5^V\epsilon^{\mu\nu\rho\sigma}(W_\mu^+\partial_\rho W^-_
    \nu-\partial_\rho W_\mu^+W^-_\nu)V_\sigma\nonumber\\
    &+&\frac{\lambda^V}{m_W^2}W_\mu^{+\nu}W_\nu^{-\rho}V_\rho^{\mu}
    +\frac{\wtil{\lambda^V}}{m_W^2}W_\mu^{+\nu}W_\nu^{-\rho}\wtil{V}_\rho^{\mu}\nonumber\\
    &+&\left.\kappa^V W_\mu^+W_\nu^-V^{\mu\nu}+\wtil{\kappa^V}W_\mu^+W_\nu^-\wtil{V}^{\mu\nu}
    \right) .
    \label{eq:Lagrangian}
\end{eqnarray}
Here $W_{\mu\nu}^\pm = \partial_\mu W_\nu^\pm - 
\partial_\nu W_\mu^\pm$, $V_{\mu\nu} = \partial_\mu V_\nu - 
\partial_\nu V_\mu$, 
$\wtil{V}^{\mu\nu}=1/2\epsilon^{\mu\nu\rho\sigma}V_{\rho\sigma}$,
and the overall coupling constants are defined as
$g_{WW\gamma}=-g\sin\theta_W$ and $g_{WWZ}=-g\cos\theta_W$, with $\theta_W$ being the weak
mixing angle. In the SM, 
$g_1^V=1$, $\kappa^V=1$ and other couplings are zero. The anomalous part in $g_1^V$,
$\kappa^V$ would be $\Delta g_1^V=g_1^V-1$, $\Delta\kappa^V=\kappa^V-1$, respectively. 
The couplings $g_1^V$, $\kappa^V$
and $\lambda^V$  are $CP$-even (both $C$ and $P$-even), 
while $g_4^V$ (odd in $C$, even in $P$), $\wtil{\kappa^V}$
and $\wtil{\lambda^V}$  (even in $C$, odd in $P$) are $CP$-odd. On the other hand 
$g_5^V$ is both $C$ and $P$-odd making it  $CP$-even. We label these sets of  $14$
anomalous couplings to be $c_i^{\cal L}$ as given in Eq.~(\ref{eq:ciL}) in appendix~\ref{apendix:a} for 
later uses.

On restricting to the  $SU(2)\times U(1)$ gauge, the coupling ($c_i^{\cal L}$) of the Lagrangian
in Eq.~(\ref{eq:Lagrangian}) can be written in terms of  the couplings of the operators
in Eq.~(\ref{eq:opertaors-dim6}) as~\cite{Hagiwara:1986vm,Hagiwara:1993ck,Wudka:1994ny,Degrande:2012wf}
\begin{eqnarray}
    \Delta g_1^Z & = & c_W\frac{m_Z^2}{2\Lambda^2} ,\nonumber\\
    g_4^V &=& g_5^V=\Delta g_1^\gamma=0 ,\nonumber\\
    \lambda^\gamma & = & \lambda^Z=\lambda^V = c_{WWW}\frac{3g^2m_W^2}{2\Lambda^2} ,\nonumber\\
    \wtil{\lambda^\gamma} & = & \wtil{\lambda^Z}=\wtil{\lambda^V} = c_{\wtil{WWW}}\frac{3g^2m_W^2}{2\Lambda^2} ,\nonumber\\
    \Delta\kappa^\gamma & = & (c_W+c_B)\frac{m_W^2}{2\Lambda^2} ,\nonumber\\
    \Delta\kappa^Z & = & (c_W-c_B\tan^2\theta_W)\frac{m_W^2}{2\Lambda^2} ,\nonumber\\
    \wtil{\kappa^\gamma} & = &
    c_{\wtil{W}}\frac{m_W^2}{2\Lambda^2} ,\nonumber\\
    \wtil{\kappa^Z} & = &
    -c_{\wtil{W}}\tan^2\theta_W\frac{m_W^2}{2\Lambda^2} .
    \label{eq:Operator-to-Lagrangian}
\end{eqnarray}
It is clear from above that some of the vertex factor couplings are dependent on each other and 
they are 
\begin{eqnarray}
    &&\Delta g_1^Z=\Delta \kappa^Z + \tan^2\theta_W \Delta \kappa^\gamma ,\nonumber\\
    &&\wtil {\kappa^Z} + \tan^2\theta_W \wtil{ \kappa^\gamma}=0 .
\end{eqnarray}
We label the non-vanishing $9$ couplings in $SU(2)\times U(1)$ gauge as $c_i^{{\cal L}_g}$
given in Eq.~(\ref{eq:ciLg}) in appendix~\ref{apendix:a} for later uses.

The anomalous $WWV$ couplings have been studied in the effective operator  approach as well
as in the effective vertex formalism subjected to  $SU(2)\times U(1)$ invariance for $e^+$-$e^-$ 
collider~\cite{Gaemers:1978hg,Hagiwara:1986vm,Bilchak:1984ur,Hagiwara:1992eh,Choudhury:1996ni,Choudhury:1999fz,Wells:2015eba,Buchalla:2013wpa,Zhang:2016zsp,Berthier:2016tkq,Bian:2015zha,Bian:2016umx}, Large Hadron electron collider (LHeC)
~\cite{Biswal:2014oaa,Cakir:2014swa,Li:2017kfk}, $e$-$\gamma$ collider~\cite{Kumar:2015lna} and
hadron collider 
(LHC)~\cite{Baur:1987mt,Dixon:1999di,Bian:2015zha,Falkowski:2016cxu,Bian:2016umx,Butter:2016cvz,Azatov:2017kzw,Baglio:2017bfe,Li:2017esm,Baglio:2018bkm,Bhatia:2018ndx,Chiesa:2018lcs,Rahaman:2019lab,Baglio:2019uty,Azatov:2019xxn}. Some $CP$-odd $WWV$ couplings have been studied
in Refs.~\cite{Choudhury:1999fz,Li:2017esm,Rahaman:2019lab}.

On the experimental side, the anomalous $WWV$ couplings have been explored and 
stringent limits on them  have been obtained 
at the LEP~\cite{Abbiendi:2000ei,Abbiendi:2003mk,Abdallah:2008sf,Schael:2013ita}, the Tevatron~\cite{Aaltonen:2007sd,Abazov:2012ze}, the  
LHC~\cite{Khachatryan:2016poo,Aad:2016ett,Aad:2016wpd,Chatrchyan:2013yaa,RebelloTeles:2013kdy,ATLAS:2012mec,Chatrchyan:2012bd,Aad:2013izg,Chatrchyan:2013fya,Aaboud:2017cgf,Sirunyan:2017bey,Aaboud:2017fye,Sirunyan:2017jej,Sirunyan:2019gkh,Sirunyan:2019dyi,Sirunyan:2019bez} and Tevatron-LHC~\cite{Corbett:2013pja}. The  tightest one-parameter limit
obtained on the  anomalous couplings from   experiments are given in 
Table~\ref{tab:aTGC_constrain_form_collider}. 
The tightest limits on operator couplings ($c_i^{\cal O}$) are obtained in 
Ref.~\cite{Sirunyan:2019gkh} for $CP$-even
ones and in Ref.~\cite{Aaboud:2017fye} for $CP$-odd ones. 
These limits translated to  $c_i^{{\cal L}_g}$ using Eq.~(\ref{eq:Operator-to-Lagrangian}) are also given in
Table~\ref{tab:aTGC_constrain_form_collider}. The tightest limits on the  couplings $g_4^Z$ and 
$g_5^Z$ are obtained  in Refs.~\cite{Abdallah:2008sf,Abbiendi:2003mk} considering the 
Lagrangian in Eq.~(\ref{eq:Lagrangian}).
\begin{table*}[!ht]
	\centering
	\caption{\label{tab:aTGC_constrain_form_collider} The list of tightest limits obtained on the
		anomalous couplings of dimension-$6$ operators in Eq~(\ref{eq:opertaors-dim6}) and
		effective vertices in Eq.~(\ref{eq:Lagrangian}) in the $SU(2)\times U(1)$ gauge (except $g_4^Z$ and $g_5^Z$) at $95\%$ C.L.  from experiments.}
	\renewcommand{\arraystretch}{1.5}
	\begin{tabular*}{\textwidth}{@{\extracolsep{\fill}}lll@{}}\hline
		$c_i^{\cal O}$            & Limits (TeV$^{-2}$)   & Remark\\\hline 
		$\frac{c_{WWW}}{\Lambda^2}$                    & $[-1.58,+1.59]$ &CMS $\sqrt{s}=13$ TeV, ${\cal L}=35.9$ fb$^{-1}$, $SU(2)\times U(1)$~\cite{Sirunyan:2019gkh} \\
		$\frac{c_{W}}{\Lambda^2}$                     & $[-2.00,+2.65]$ &CMS~\cite{Sirunyan:2019gkh} \\
		$\frac{c_{B}}{\Lambda^2}$                    & $[-8.78,+8.54]$ &CMS~\cite{Sirunyan:2019gkh} \\    
		$ \frac{c_{\widetilde{WWW}}}{\Lambda^2}$    &$[-11,+11]$  &ATLAS $\sqrt{s}=7(8)$ TeV, ${\cal L}=4.7(20.2)$ fb$^{-1}$ ~\cite{Aaboud:2017fye}\\
		$ \frac{c_{\widetilde{W}}}{\Lambda^2}$      &$[-580,580]$  &ATLAS~\cite{Aaboud:2017fye} \\
		\hline
		$c_i^{{\cal L}_g}$ & Limits ($\times 10^{-2}$) & Remark\\ \hline
		$\lambda^V$ &  $[-0.65,+0.66]$ &CMS~\cite{Sirunyan:2019gkh}\\
		$\Delta\kappa^\gamma$ &$[-4.4,+6.3]$&CMS $\sqrt{s}=8$ TeV, ${\cal L}=19$ fb$^{-1}$, $SU(2)\times U(1)$~\cite{Sirunyan:2017bey}\\
		$\Delta g_1^Z$ & $[-0.61,+0.74]$ &  CMS~\cite{Sirunyan:2019gkh}\\
		$\Delta\kappa^Z$ & $[-0.79,+0.82]$ &CMS~\cite{Sirunyan:2019gkh}\\
		$\wtil{\lambda^V}$ & $[-4.7,+4.6]$ &ATLAS~\cite{Aaboud:2017fye}\\
		$\wtil{\kappa^Z}$  & $[-14,-1]$ & DELPHI (LEP2), $\sqrt{s}=189$-$209$ GeV, ${\cal L}=520$ pb$^{-1}$~\cite{Abdallah:2008sf}\\
		\hline
		$c_i^{{\cal L}}$  & Limits ($\times 10^{-2}$) & Remark\\ \hline
		$g_4^Z$ & $[-59,-20]$ &DELPHI~\cite{Abdallah:2008sf}\\ 
		$g_5^Z$  & $[-16,+9.0]$ &OPAL (LEP), $\sqrt{s}=183$-$209$ GeV, ${\cal L}=680$ pb$^{-1}$~\cite{Abbiendi:2003mk} \\
		\hline
	\end{tabular*}
\end{table*}

The  $W^+W^-$ production is one of the important processes to be studied
at the future International Linear Collider (ILC)~\cite{Djouadi:2007ik,Baer:2013cma,Behnke:2013xla} for the precision test~\cite{MoortgatPick:2005cw} as well as 
for BSM physics. This process has been studied earlier for SM phenomenology
as well as for various BSM physics with and without beam 
polarization~\cite{Hagiwara:1986vm,Gounaris:1992kp,Ananthanarayan:2009dw,Ananthanarayan:2010bt,Ananthanarayan:2011ga,Andreev:2012cj}.
Here we intend to study $WWV$ anomalous couplings in $e^+e^-\to W^+W^-$ at $\sqrt{s}=500$ GeV  and integrated luminosity of 
${\cal L}=100$ fb$^{-1}$ using the cross section, forward-backward asymmetry, and eight
polarizations asymmetries  of $W^-$ for a set of choices of longitudinally  polarized $e^+$ and $e^-$ beams
in the channel $W^- \to l^- \bar{\nu_l}$ ($l=e,\mu$)\footnote{For simplicity we do not include the  tau decay mode as the tau decays to the neutrino within the beam pipe,  
    giving extra missing momenta affecting the reconstruction of the events.} and $W^+\to hadrons$.
The polarizations of $Z$ and $W$ are being used widely recently for various BSM studies~\cite{Aguilar-Saavedra:2017zkn,Renard:2018tae,Renard:2018bsp,Renard:2018lqv,Renard:2018jxe,Renard:2018blr,Behera:2018ryv} along with studies with anomalous gauge boson couplings~\cite{Abbiendi:2000ei,Rahaman:2016pqj,Rahaman:2017qql,Rahaman:2018ujg}. Recently the polarizations of $W/Z$ have been measured in $WZ$ production at the LHC~\cite{Aaboud:2019gxl}.
Besides the final state polarizations, the initial state beam polarizations at the ILC can be used to enhance the relevant signal to background ratio~\cite{MoortgatPick:2005cw,Pankov:2005kd,Osland:2009dp,Ananthanarayan:2010bt,Andreev:2012cj}. It also has the ability to distinguish between $CP$-even and  $CP$-odd couplings~\cite{Choudhury:1994nt,Czyz:1988yt,Ananthanarayan:2003wi,Ananthanarayan:2004eb,MoortgatPick:2005cw,Bartl:2005uh,Rao:2006hn,Bartl:2007qy,Dreiner:2010ib,Kittel:2011rk,Ananthanarayan:2011fr}.
 We note that  an 
$e^+e^-$ machine will run with longitudinal beam polarizations switching between $(\eta_3,\xi_3)$ and $(-\eta_3,-\xi_3)$~\cite{MoortgatPick:2005cw},
where $\eta_3(\xi_3$) is the longitudinal polarization of $e^-$ ( $e^+$).
For an integrated luminosity of  $100$ fb$^{-1}$, one will have half the luminosity available for each polarization
configuration. The most common observables, the cross section for example, 
studied in literature with beam polarizations are the total cross section
\begin{equation}\label{eq:sigma_T}
\sigma_T(\eta_3,\xi_3)=\sigma(+\eta_3,+\xi_3)+\sigma(-\eta_3,-\xi_3)
\end{equation}
and the difference
\begin{equation}\label{eq:sigma_A}
\sigma_A(\eta_3,\xi_3)=\sigma(+\eta_3,+\xi_3)-\sigma(-\eta_3,-\xi_3) .
\end{equation}
We find that  combining the two opposite beam polarizations at the level of $\chi^2$ rather than combining them as  in Eqs.~(\ref{eq:sigma_T}) \&~(\ref{eq:sigma_A}), we can constrain the anomalous 
couplings better  in this analysis; see appendix~\ref{appendix:c} for explanation. 

We note that there exist $64$ polarization correlations~\cite{Hagiwara:1986vm} 
apart from $8+8$ polarizations for $W^+$ and $W^-$. The measurement of these 
correlations requires the identification of light quark falvors in the
above channel, which is not possible; hence,  we are not including polarization correlations in our analysis.
In the case of both the $W$s decaying leptonicaly, there are two  missing neutrinos and  
reconstruction of polarization observables suffers combinatorial ambiguity. 
Here we aim to work with a set of observables that can be reconstructed 
uniquely and test their ability to probe the anomalous couplings including partial contribution
up to ${\cal O}(\Lambda^{-4})$\footnote[2]{ We calculate the  cross section
    up to ${\cal O}( 
    \Lambda^{-4})$, i.e., quadratic in dimension-$6$ (as linear approximation is not valid; see appendix~\ref{appendix:b})  and linear in dimension-$8$ couplings choosing dimension-$8$ couplings to be zero to
    compare our result with current LHC constraints on
    dimension-$6$ parameters~\cite{Sirunyan:2019gkh,Aaboud:2017fye}.}.

The rest of the paper is arranged in the following way. In Sect.~\ref{sec:2} we introduce 
the complete set polarization observables of a spin-$1$ particle along with 
the forward-backward asymmetry and study the effect of beam polarizations on the observables.
In Sect.~\ref{sec:3} we use the vertex form factors for the Lagrangian in Eq.~(\ref{eq:Lagrangian})
and obtain expressions for all the observables. In this section, 
we cross-validate analytical results against the numerical result from 
{\tt MadGraph5}~\cite{Alwall:2014hca} for sanity checking. We also study the  $\cos\theta$ (of $W$)
dependences  of the observables  and study their sensitivity on the anomalous couplings. 
In this section, we also estimate simultaneous limits on $c_i^{{\cal L}}$,   
$c_i^{\cal O}$  and the translated limits on
$c_i^{{\cal L}_g}$.   We give an insight 
into the choice of beam polarizations in this process in Sect.~\ref{sec:3.3} and conclude in Sect.~\ref{sec:conclusion}.
\section{Observables and effect of beam polarizations}\label{sec:2}
\begin{figure}[ht!]
    \centering
    \includegraphics[width=0.496\textwidth]{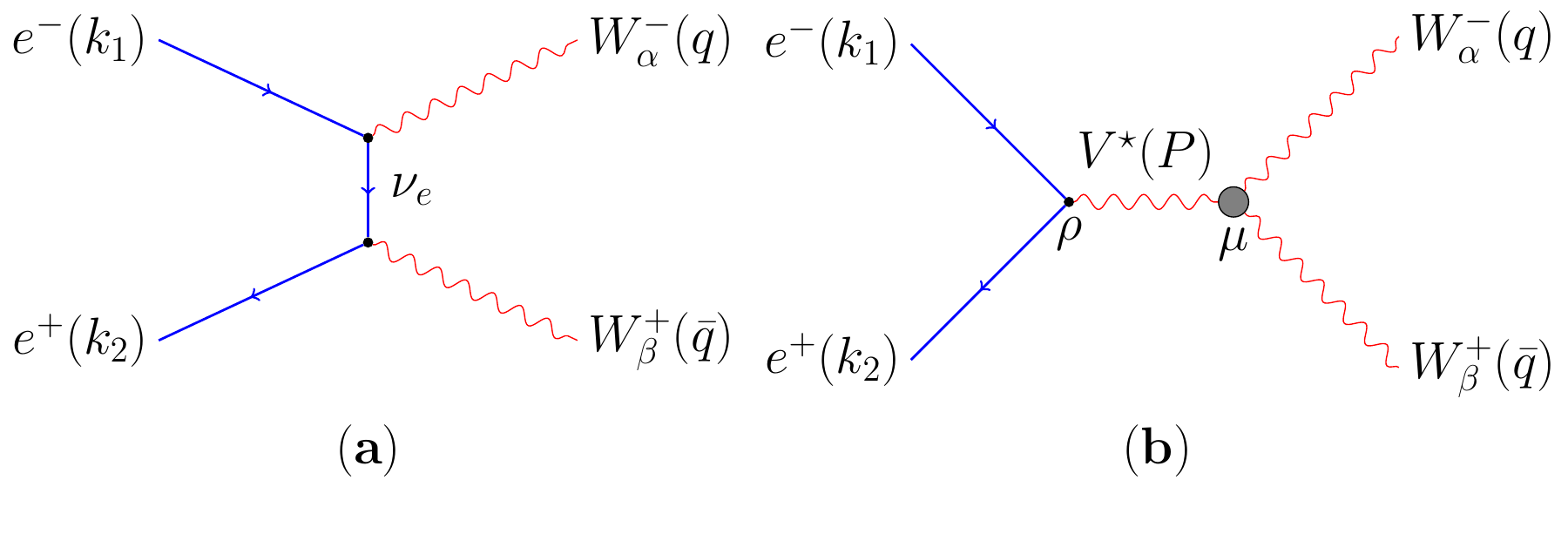}
    \caption{\label{fig:Feynman-ee-ww}Feynman diagrams of $e^+e^-\to W^+W^-$,
        (a) $t$-channel and (b) $s$-channel with anomalous $W^+W^-V$ 
        ($V=\gamma,Z$) vertex contribution shown by shaded blob.} 
\end{figure}
\begin{figure*}[ht!]
	\centering
	\includegraphics[width=0.485\textwidth]{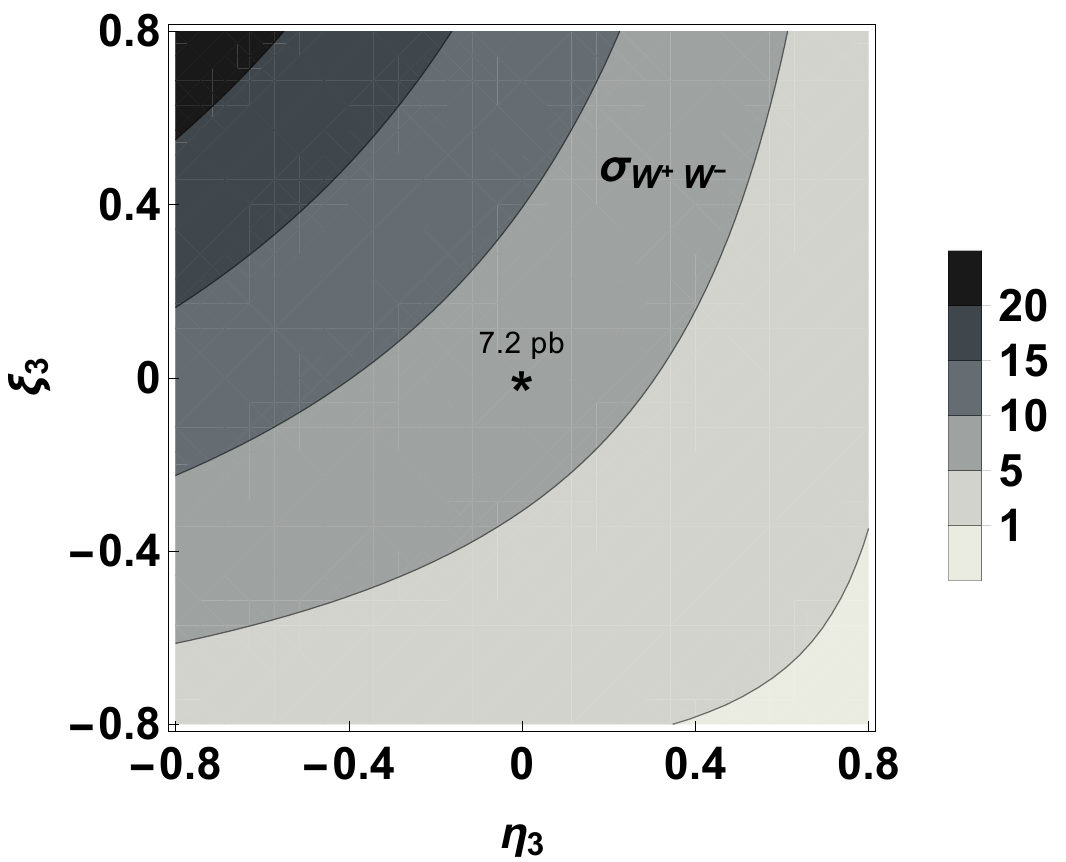}
	\includegraphics[width=0.506\textwidth]{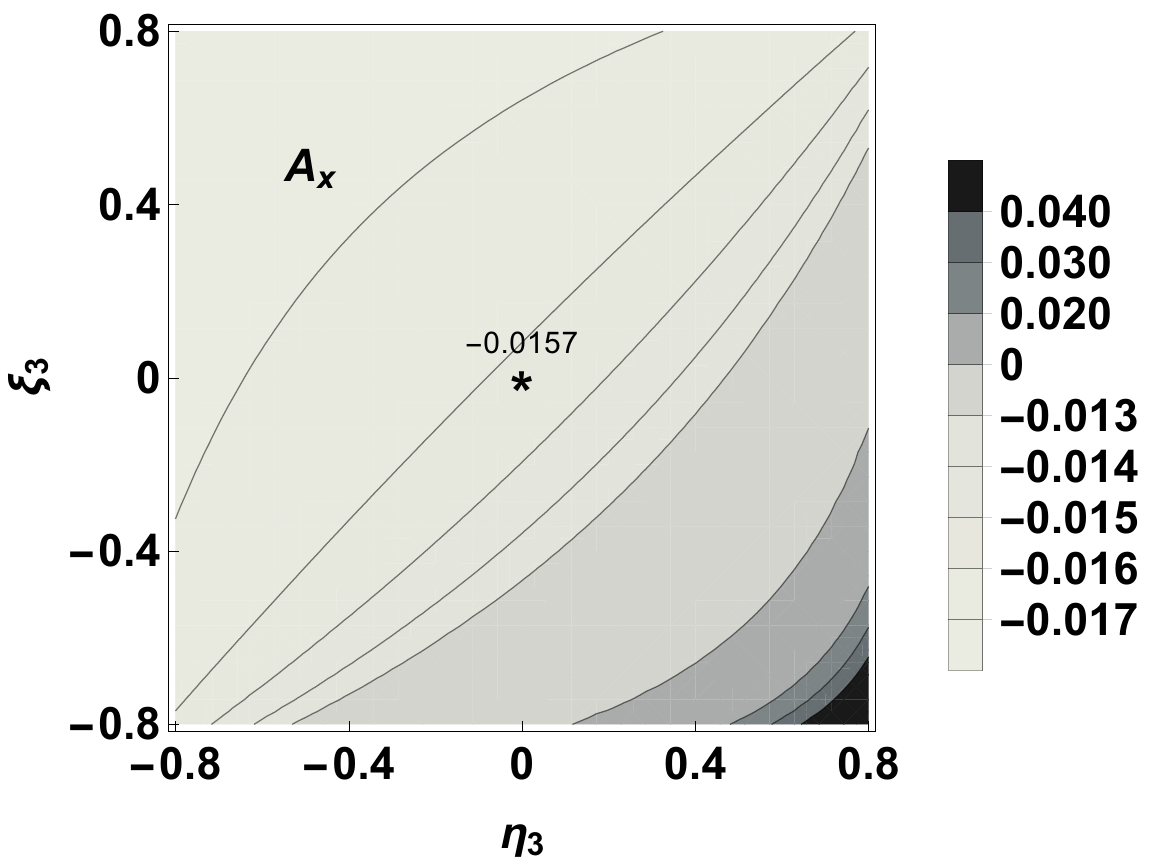}
	\caption{\label{fig:Sigma_and_Afb_eta3xi3}  The production cross section $\sigma_{W^+W^-}$ in pb ({\em left-panel})
		and the polarization asymmetry $A_x$ ({\em right-panel}) in the SM as a function of longitudinal beam polarizations 
		$\eta_3$ (for $e^-$) and $\xi_3$ (for $e^+$)   at $\sqrt{s}=500$ GeV. 
		The asterisks represent the unpolarized points and the numbers near them correspond to the SM
		values for corresponding observables with unpolarized beams.} 
\end{figure*}
We study $W^+W^-$ production at ILC running at $\sqrt{s}=500$ GeV and 
integrated luminosity ${\cal L}=100$ fb$^{-1}$ using longitudinal polarization  
of $e^-$ and $e^+$ beams giving $50$ fb$^{-1}$ to each choice of beam polarization.
The Feynman diagrams for the process are  shown in 
Fig.~\ref{fig:Feynman-ee-ww} where Fig.~\ref{fig:Feynman-ee-ww}(\textbf{a}) corresponds
to the $\nu_e$ mediated  $t$-channel diagram and the 
Fig.~\ref{fig:Feynman-ee-ww}(\textbf{b}) corresponds to the $V~(Z/\gamma)$ mediated $s$-channel 
diagram containing the anomalous triple gauge boson couplings (aTGC) contributions represented by the shaded blob. The decay mode is chosen to be
\begin{equation}
W^+\to q_u~\bar{q}_d \ \   ,~~~~~  W^-\to l^-~\bar{\nu}_l , ~~~l=e,\mu ,
\end{equation}
where $q_u$ and $q_d$ are up-type and down-type quarks, respectively.
We use complete set of eight spin-$1$  observables of $W^-$ boson
~\cite{Aguilar-Saavedra:2015yza,Rahaman:2016pqj}.

The $W$ boson being a spin-$1$ particle, its normalized production density matrix
in the spin basis can be written as~\cite{Bourrely:1980mr,Boudjema:2009fz}
\begin{equation}\label{eq:spin-desnity-matrix}
\rho(\lambda,\lambda^\prime)=\dfrac{1}{3}\Bigg[I_{3\times 3} +\dfrac{3}{2} \vec{p}.\vec{S}
+\sqrt{\dfrac{3}{2}} T_{ij}\big(S_iS_j+S_jS_i\big) \Bigg] ,
\end{equation}
where $\vec{p}=\{p_x,p_y,p_z\}$ is the vector polarization of a spin-$1$ particle,
$\vec{S}=\{S_x,S_y,S_z\}$ is the spin basis and $T_{ij}~(i,j=x,y,z)$
is the $2^{nd}$-rank symmetric traceless tensor, and $\lambda$ and $\lambda^\prime$ are
helicities of the particle. The tensor $T_{ij}$ has five independent elements, which are
$T_{xy}$, $T_{xz}$, $T_{yz}$, $T_{xx}-T_{yy}$ and $T_{zz }$. 
Combining the $\rho(\lambda,\lambda^\prime)$ with the   normalized
decay density matrix of the  particle  to a pair of fermion $f$, 
the differential cross section  would be~\cite{Boudjema:2009fz}
\begin{eqnarray}
\frac{1}{\sigma} \ \frac{d\sigma}{d\Omega_f} &=&\frac{3}{8\pi} \left[
\left(\frac{2}{3}-(1-3\delta) \ \frac{T_{zz}}{\sqrt{6}}\right) + \alpha \ p_z
\cos\theta_f \right.\nonumber\\
&+& \sqrt{\frac{3}{2}}(1-3\delta) \ T_{zz} \cos^2\theta_f 
\nonumber\\
&+&\left(\alpha \ p_x + 2\sqrt{\frac{2}{3}} (1-3\delta)
\ T_{xz} \cos\theta_f\right) \sin\theta_f \ \cos\phi_f \nonumber\\
&+&\left(\alpha \ p_y + 2\sqrt{\frac{2}{3}} (1-3\delta)
\ T_{yz} \cos\theta_f\right) \sin\theta_f \ \sin\phi_f \nonumber\\
&+&(1-3\delta) \left(\frac{T_{xx}-T_{yy}}{\sqrt{6}} \right) \sin^2\theta_f
\cos(2\phi_f)\nonumber\\
&+&\left. \sqrt{\frac{2}{3}}(1-3\delta) \ T_{xy} \ \sin^2\theta_f \
\sin(2\phi_f) \right] .
\label{eq:angular_distribution}
\end{eqnarray}
Here $\theta_f$, $\phi_f$ are the polar and the azimuthal orientation of the  fermion $f$,
in the rest frame of the particle ($W$) with its would-be momentum along the $z$-direction. 
The initial beam direction and the $W^-$ momentum in the lab frame  define   the $x$--$z$ plane, i.e. $\phi = 0$ plane, in the rest frame of $W^-$ as well.
In this case
$\alpha=-1$ and $\delta=0$. The vector polarizations $\vec{p}$ and independent tensor
polarizations $T_{ij}$ are calculable from the asymmetries constructed from the 
decay angular distribution of the lepton (here $l^-$). For example 
$p_x$ can be calculated from the asymmetry $A_x$ as 
\begin{eqnarray}\label{eq:pol_decay_Ax}
A_x=
\dfrac{\sigma(\cos\phi_f>0)-\sigma(\cos\phi_f<0)}{\sigma(\cos\phi_f>0)+\sigma(\cos\phi_f<0)}
\equiv   \frac{3 \alpha  p_x}{4} .
\end{eqnarray}
The asymmetries corresponding to all other polarizations,  vector polarizations  
$p_y$, $p_z$ and independent tensor polarizations $T_{ij}$
are $A_y$, $A_z$, $A_{xy}$, $A_{xz}$, $A_{yz}$, $A_{x^2-y^2}$, $A_{zz}$; see Ref.~\cite{Rahaman:2016pqj} for details. 

Owing to the 
$t$-channel process (Fig.~\ref{fig:Feynman-ee-ww}\textbf{a}) and absence of a $u$-channel 
process,  like in $ZV$ production~\cite{Rahaman:2016pqj,Rahaman:2017qql},
the $W^\pm$ produced are not 
forward-backward symmetric. We include the forward-backward asymmetry of $W^-$, defined as 
\begin{equation}
A_{fb}=\frac{1}{\sigma_{W^+W^-}}\Bigg[\int_0^1 \frac{d\sigma_{W^+W^-}}{d\cos\theta_{W^-}} 
-\int_{-1}^0 \frac{d\sigma_{W^+W^-}}{d\cos\theta_{W^-}}    \Bigg] ,
\end{equation}
to the set of observables making a total of ten observables including
the cross section as well. Here $\theta_{W^-}$ is the production angle of 
the $W^-$ with respect to the $e^-$ beam direction and $\sigma_{W^+W^-}$ is the production cross 
section. 

These asymmetries can be measured in a real collider
from the final state lepton $l^-$. One has to calculate the asymmetries 
in the rest frame of $W^-$ which require the missing $\bar{\nu_l}$ momenta to be
reconstructed. At an $e^+$ $e^-$ collider, as studied here, reconstructing the missing 
$\bar{\nu_l}$ is possible because only one missing particle is involved and
no parton distribution functions (PDFs)
are involved, i.e., initial momenta are known. But 
for a collider where PDFs are involved, reconstructing the actual missing momenta
may not be  possible.

We explore the dependence of the cross section and asymmetries on the
longitudinal polarization $\eta_3$ of $e^-$ and $\xi_3$ of $e^+$.   
In Fig.~\ref{fig:Sigma_and_Afb_eta3xi3}, we show the production cross section 
$\sigma_{W^+W^-}$  and $A_x$ as a function of beam polarizations as an example. 
The cross section decreases along the $\eta_3=-\xi_3$ path from $20$ pb on the 
left-top corner to $7.2$ pb at the unpolarized point and further to $1$ pb in the 
right-bottom corner. This is because of  the $W^\pm$ 
couples to the left chiral $e^-$ i.e., it requires $e^-$ to be negatively 
polarized and $e^+$ to be positively polarized for the higher cross section.
The variation 
of $A_{fb}$ (not shown) with the beam polarization is the same as the cross section but 
 very slow above the line $\eta_3=\xi_3$. From this, we can expect 
that a positive $\eta_3$ and a negative $\xi_3$ will reduce the SM contributions to  
observables increasing the $S/\sqrt{B}$ ratio ($S=$ signal, $B=$ background).
Some other asymmetries, like $A_x$, have the opposite dependence on the beam
polarizations compared to the cross section; their modulus reduce for negative $\eta_3$ and positive
$\xi_3$.
\section{Probe to the anomalous couplings}\label{sec:3}
\begin{figure}[h!]
    \centering
    \includegraphics[width=0.496\textwidth]{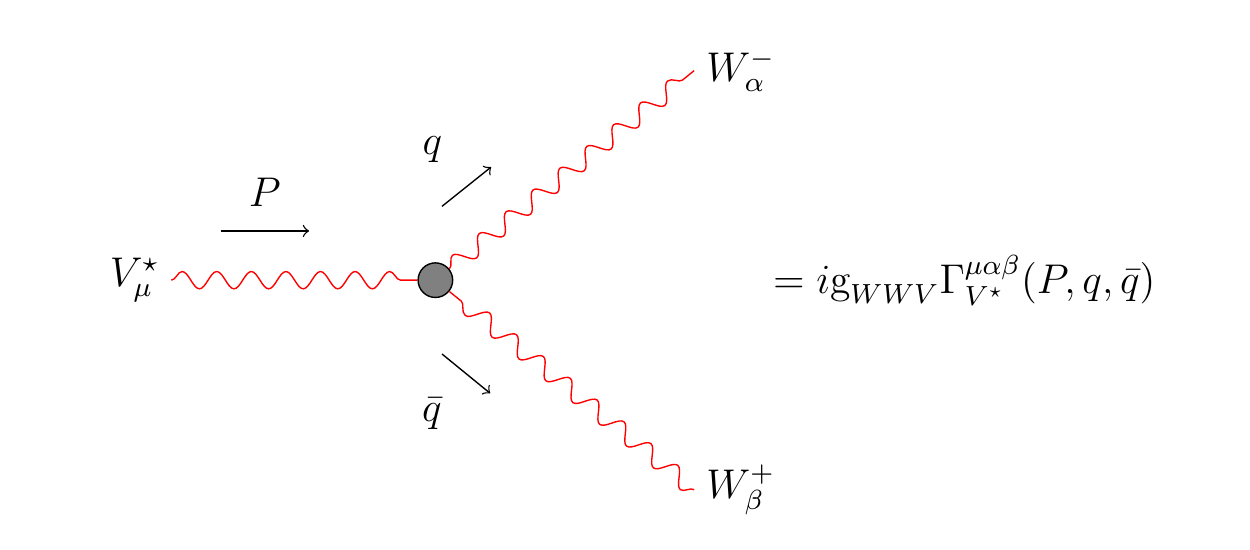}
    \caption{\label{fig:wwv_vertex}The $WWV$ vertex showing anomalous contribution 
        represented by the  shaded blob on top of SM. The momentum $P$ is incoming to the vertex, 
        while $q$ and $\bar{q}$ are outgoing from the vertex.} 
\end{figure}
\begin{figure*}[htb!]
	\centering
	\includegraphics[width=0.496\textwidth]{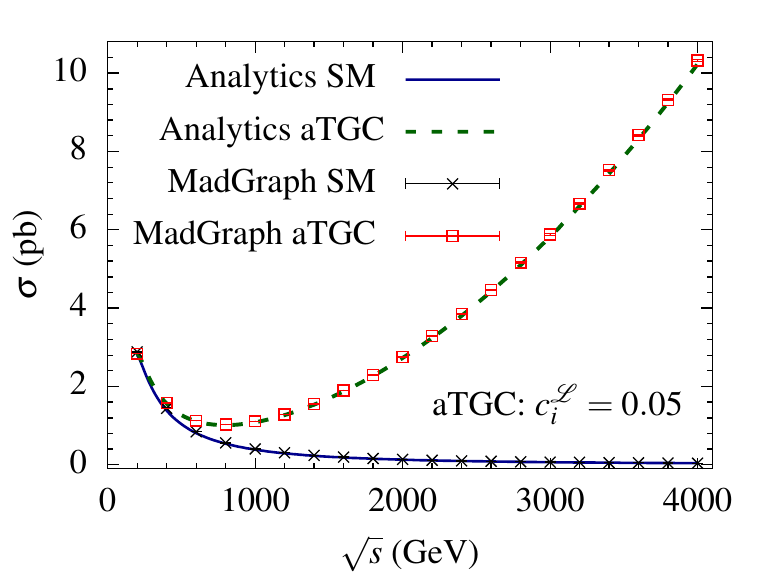}
	\includegraphics[width=0.496\textwidth]{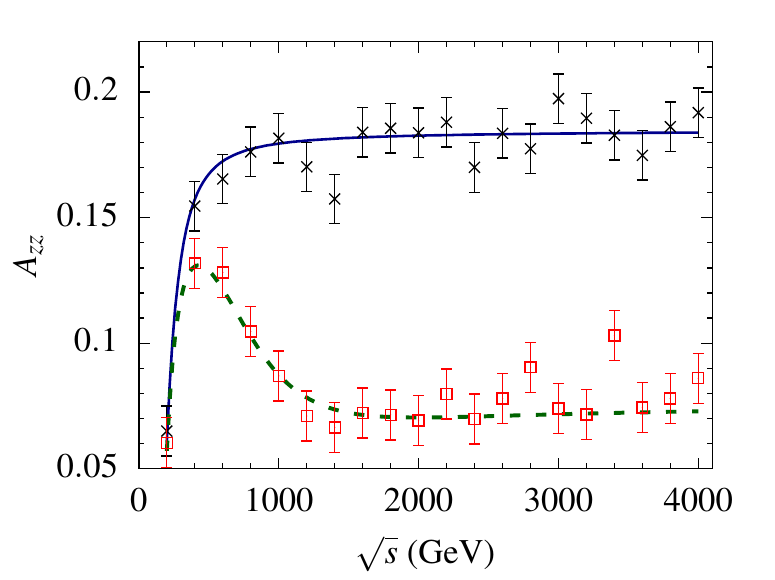}
	\caption{\label{fig:SanityCheck}  The cross section $\sigma$ including the decays in pb ({\em left-panel})
		and the  asymmetry $A_{zz}$ ({\em right-panel}) in the SM   and {\tt aTGC} with all anomalous 
		couplings  ($c_i^{\cal L}$) at $0.05$   
		as a function of $\sqrt{s}$  for  the SM analytic ({\it solid}/blue) and {\tt aTGC} analytic
		({\it dashed} /green) with unpolarized beams. The {\it crossed} (black) points and {\it boxed} (red) 
		points with the error bar correspond to results from {\tt MadGraph5}.
		The error bars are given for number of events of $10^4$. } 
\end{figure*}
The $W^+W^-V$ vertex (Fig.~\ref{fig:wwv_vertex})
for the Lagrangian in Eq.~(\ref{eq:Lagrangian}) for on-shell $W$s would be 
$ig_{WWV}\Gamma_V^{\mu\alpha\beta}$ \cite{Gaemers:1978hg,Hagiwara:1986vm}
and it is given by
\begin{eqnarray}
\Gamma_V^{\mu\alpha\beta}&=&f_1^V(q-\bar q)^\mu g^{\alpha\beta}-
\frac{f_2^V}{m_W^2}(q-\bar q)^\mu P^\alpha P^\beta\nonumber\\&&
+f_3^V(P^\alpha g^{\mu\beta}-P^\beta g^{\mu\alpha})\nonumber
+if_4^V(P^\alpha g^{\mu\beta}+P^\beta g^{\mu\alpha})\nonumber\\&&
+if_5^V\epsilon^{\mu\alpha\beta\rho}(q-\bar{q})_\rho
-f_6^V\epsilon^{\mu\alpha\beta\rho}P_\rho \nonumber\\&&
+\frac{\wtil{f_7^V}}{m_W^2}
\left(\bar{q}^\alpha\epsilon^{\mu\beta\rho\sigma} + 
q^\beta\epsilon^{\mu\alpha\rho\sigma}\right)q_\rho\bar{q}_\sigma,
\label{eq:wwv_vertex}
\end{eqnarray}
where $P,q,\bar q$ are the four-momenta of $V,W^-,W^+$, respectively. The 
momentum conventions are shown in  Fig.~\ref{fig:wwv_vertex}. 
The form factors $f_i$s have been  obtained from the Lagrangian in Eq.~(\ref{eq:Lagrangian}) using  
{\tt FeynRules}~\cite{Alloul:2013bka} to be
\begin{eqnarray}\label{eq:reltn_f_Lagrn}
&&f_1^V=g_1^V + \frac{\hat{s}}{2m_W^2}\lambda^V, \hspace{0.2cm}
f_2^V=\lambda^V,\hspace{0.2cm}
f_3^V=g_1^V + \kappa^V + \lambda^V , \nonumber\\
&&f_4^V=g_4^V,\hspace{0.2cm}
f_5^V=g_5^V,\hspace{0.2cm}
f_6^V=\widetilde{\kappa^V} +
\left(1-\frac{\hat{s}}{2m_W^2} \right)\widetilde{\lambda^V}, \nonumber\\
&&\wtil{f_7^V}=\widetilde{\lambda^V}.
\end{eqnarray}
We use the vertex factors in Eq.~(\ref{eq:wwv_vertex}) for the analytical 
calculation of our observables and cross validate them  numerically
with {\tt MadGraph5}~\cite{Alwall:2014hca} implementation
of Eq.~(\ref{eq:Lagrangian}). As an example, we present two observables 
$\sigma_{W^+W^-}$ and $A_{zz}$ for the SM ($c_i^{\cal L}=0.0$) and for a 
chosen couplings point $c_i^{\cal L}=0.05$, in Fig.~\ref{fig:SanityCheck}. 
The agreement between the analytical and the numerical calculations over a range
of $\sqrt{s}$ indicates the validity of relations in Eq.~(\ref{eq:reltn_f_Lagrn}),
especially the $s$ dependence of $f_1^V$ and $f_6^V$.

Analytical  expressions of all the observables have been obtained and their
dependence on the anomalous couplings $c_i^{\cal L}$ are given  in
Table~\ref{tab:param_dependence} in appendix~\ref{apendix:a}.
The $CP$-even couplings in $CP$-even observables $\sigma$, $A_x$, $A_z$, $A_{xz}$
, $A_{x^2-y^2}$, and $A_{zz}$ appear in linear as well as in quadratic form but do not
appear  in the $CP$-odd observables $A_y$, $A_{xy}$, and $A_{yz}$. On the other hand, 
$CP$-odd couplings appear linearly in  $CP$-odd observables and quadratically
in $CP$-even observables. Thus the $CP$-even couplings may have
a double patch in their confidence interval leading to asymmetric limits  
which will be discussed in Sect.~\ref{sec:3.1}. On the other hand, the $CP$-odd
couplings will have a single patch in their confidence interval and will pose 
symmetric limits. 
\subsection{Sensitivity of observables on anomalous couplings and their binning}
\label{sec:3.1}
\begin{figure*}
	\centering
	\includegraphics[width=0.496\textwidth]{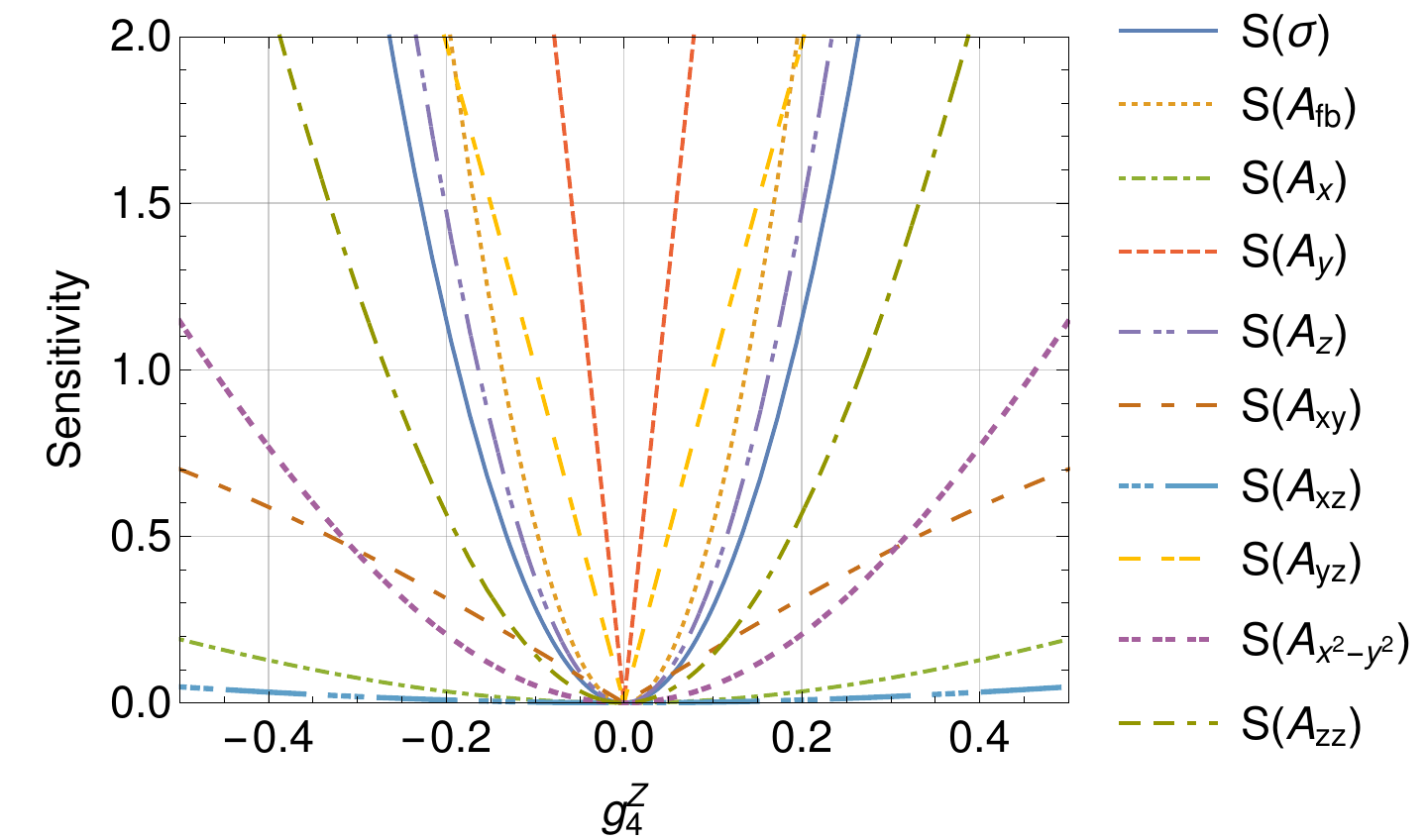}
	\includegraphics[width=0.496\textwidth]{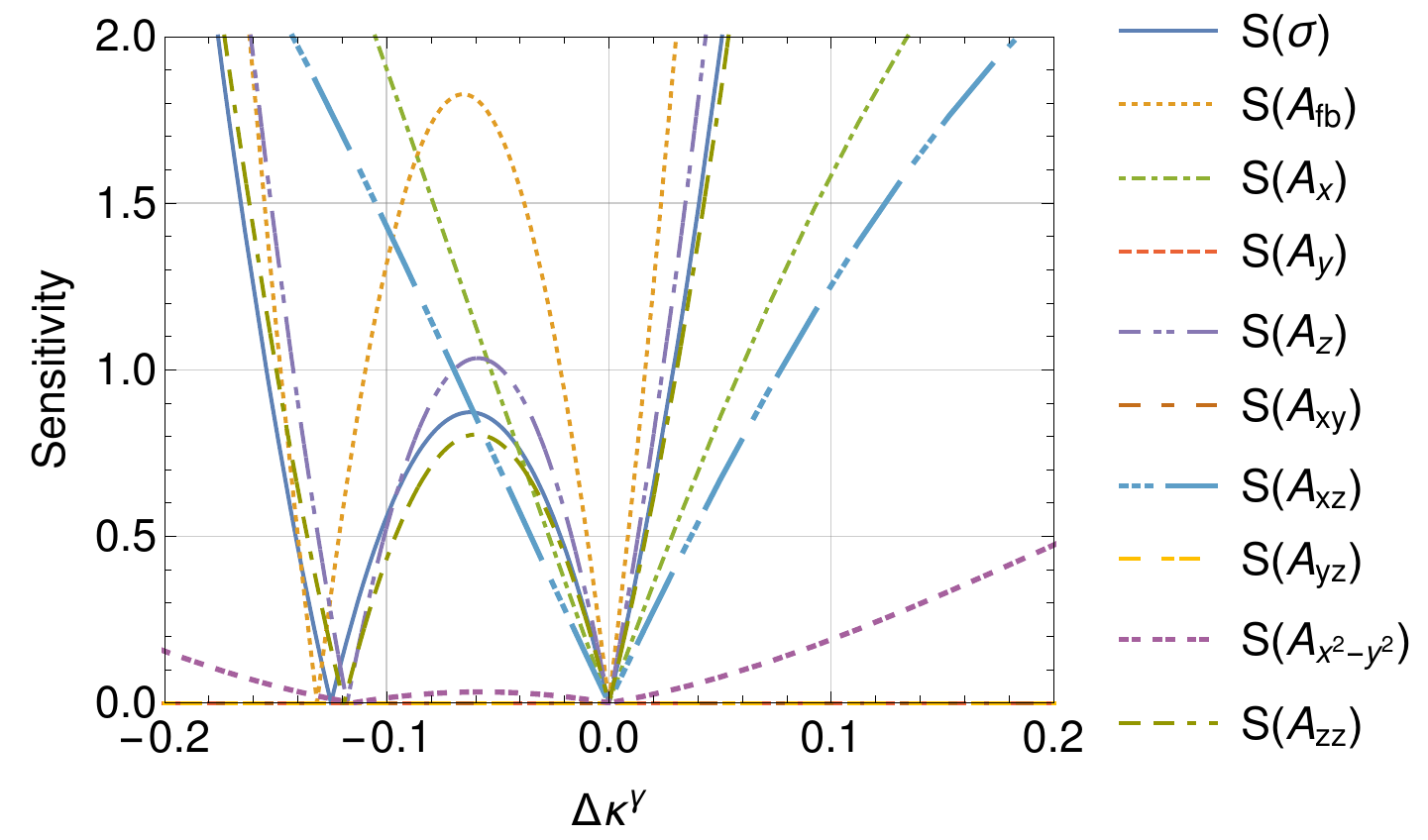}
	\caption{\label{fig:sensitivity} The one-parameter sensitivities of  cross section $\sigma$, 
		$A_{fb}$ and the eight polarization asymmetries ($A_i$) on $g_4^Z$ ({\em left-panel})
		and on $\Delta\kappa^\gamma$ ({\em right-panel})   for $\sqrt{s}=500$ GeV, ${\cal L}=100$ fb$^{-1}$ with unpolarized beams.} 
\end{figure*}
The sensitivity of an observable ${\cal O}$ depending on anomalous couplings 
$\vec{f}$ with beam polarization $\eta_3$, $\xi_3$ is given by
\begin{equation}
S{\cal O}(\vec{f},\eta_3,\xi_3)=\dfrac{|{\cal O}(\vec{f},\eta_3,\xi_3)
    -{\cal O}(\vec{0},\eta_3, \xi_3)|}{|\delta{\cal O}(\eta_3, \xi_3)|} ,
\end{equation}
where $\delta{\cal O}=\sqrt{(\delta{\cal O}_{stat.})^2+
    (\delta{\cal O}_{sys.})^2}$ is the estimated error in ${\cal O}$. 
The  error for the cross section would be,
\begin{equation}
\delta\sigma(\eta_3, \xi_3)=\sqrt{\frac{\sigma(\eta_3, \xi_3)}{{\cal L}} +
    \epsilon_\sigma^2 \sigma(\eta_3, \xi_3)^2 }
\end{equation}
whereas the estimated error in the asymmetries would be,
\begin{equation}
\delta A(\eta_3, \xi_3)=\sqrt{\frac{1-A(\eta_3, \xi_3)^2}
    {{\cal L}\sigma(\eta_3, \xi_3)} + \epsilon_A^2 } .
\end{equation}
Here ${\cal L}$ is the  luminosity of the data set, $\epsilon_\sigma$, and 
$\epsilon_A$ are the systematic fractional errors in the cross section and 
asymmetries, respectively. We take ${\cal L}=50$ fb$^{-1}$ for each choice of beam polarizations,
$\epsilon_\sigma=2~\%$ and $\epsilon_A=1~\%$, as a benchmark scenario for the present
analyses. 
The sensitivity of all $10$ observables have been studied on  all  $14$
couplings of the Lagrangian in Eq.~(\ref{eq:Lagrangian}) with the chosen $\sqrt{s}$,
${\cal L}$ and systematic uncertainties. The sensitivity of all observables 
on $g_4^Z$ and $\Delta\kappa^\gamma$ are shown in Fig.~\ref{fig:sensitivity}
as representative. Being $CP$-odd (either only linear or only quadratic terms
present), $g_4^Z$ has a single patch in the confidence interval, 
while the  $\Delta\kappa^\gamma$ being  $CP$-even 
(linear as well as quadratic terms present), 
 has two patches in the sensitivity curve, as noted earlier. The  $CP$-odd observable $A_y$
provides the tightest one-parameter limit on $g_4^Z$. The tightest $1\sigma$ 
limit on $\Delta\kappa^\gamma$ is obtained using $A_{fb}$, while at $2\sigma$ 
level, a combination of $A_{fb}$ and $A_x$ provide the tightest limit.

\begin{figure*}
    \centering
    \includegraphics[width=0.49\textwidth]{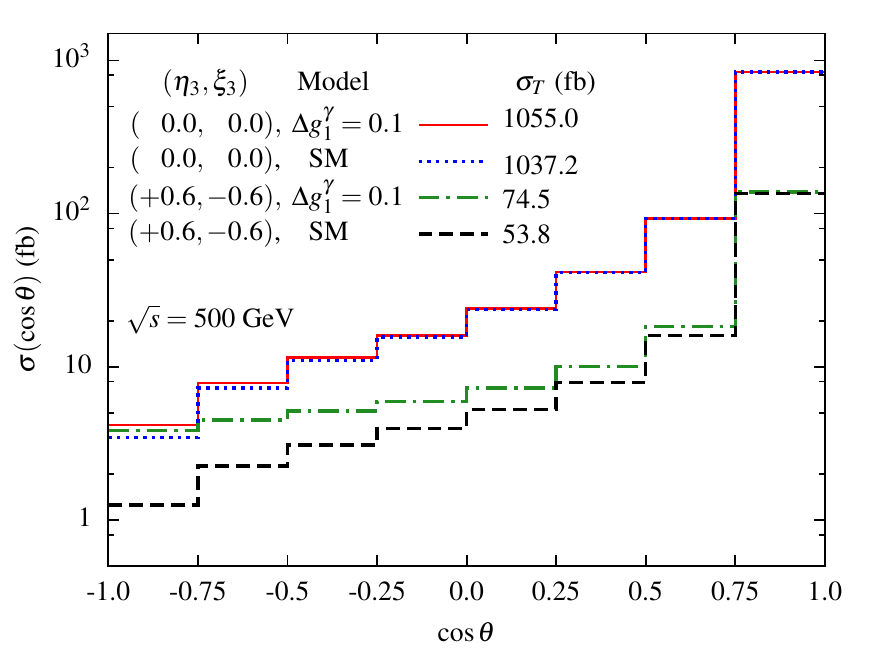}
    \includegraphics[width=0.49\textwidth]{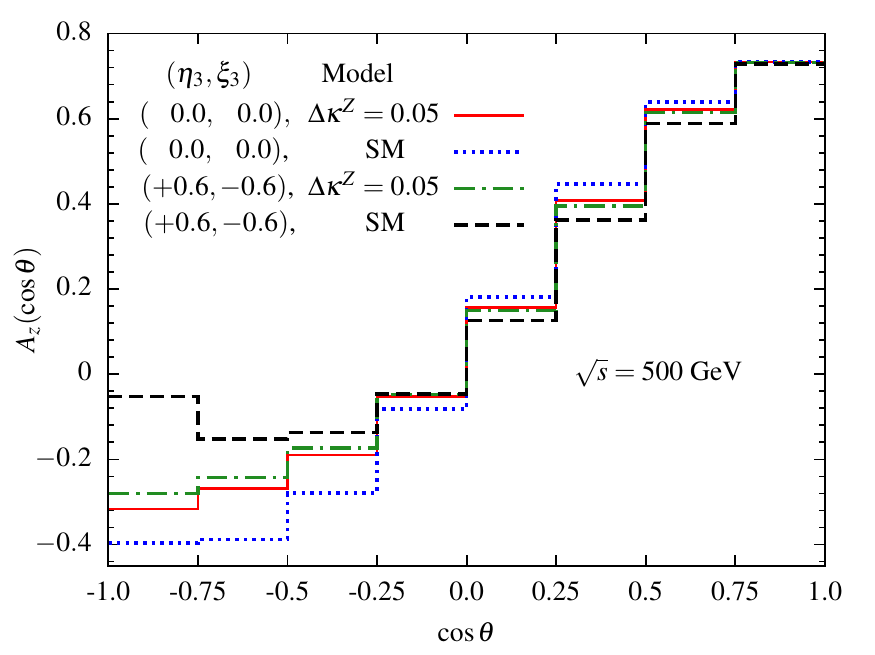}
    \includegraphics[width=0.49\textwidth]{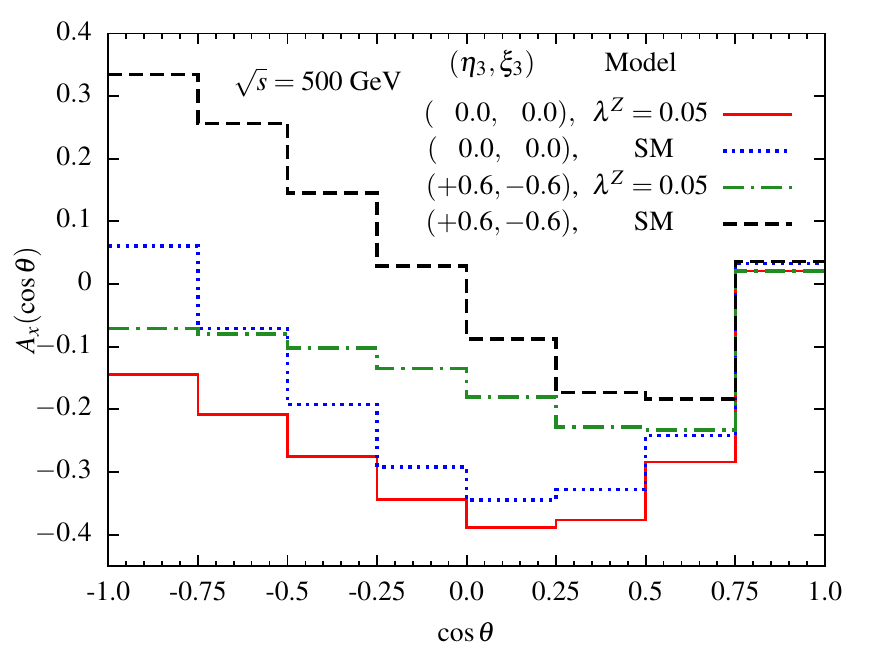}
    \includegraphics[width=0.49\textwidth]{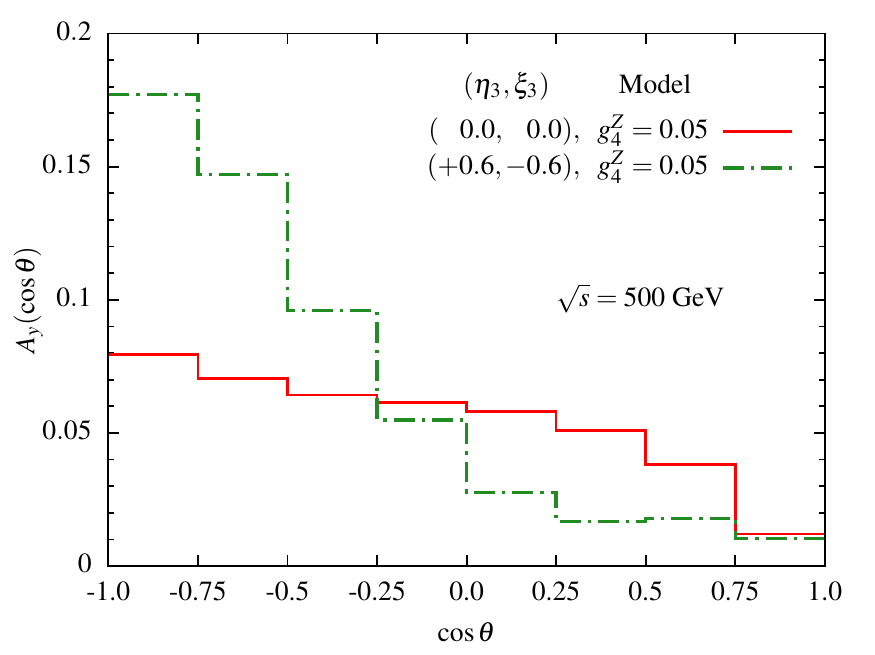}
    \caption{\label{fig:diffSigmaSM2} The cross section  $\sigma$  ({\em left-top}), $A_z$
        ({\em right-top}), $A_x$ ({\em left-bottom}) and $A_y$ ({\em right-bottom}) as a function of $\cos\theta$  
        of $W^-$ in $8$  bins  for $\sqrt{s}=500$ GeV. The {\it dotted} (blue) lines correspond to the SM unpolarized values, 
        {\it solid} (red) lines correspond to the
        unpolarized {\tt aTGC} values, {\it dashed} (black) 
        lines represent polarized SM values, and {\it dashed-dotted} (green) lines 
        represent polarized {\tt aTGC} values of 
        observables. For {\tt aTGC}, only one anomalous coupling  
        has been assumed nonzero and others kept at zero in each {\em panel}.} 
\end{figure*}
\begin{figure*}
	\centering
	\includegraphics[width=0.49\textwidth]{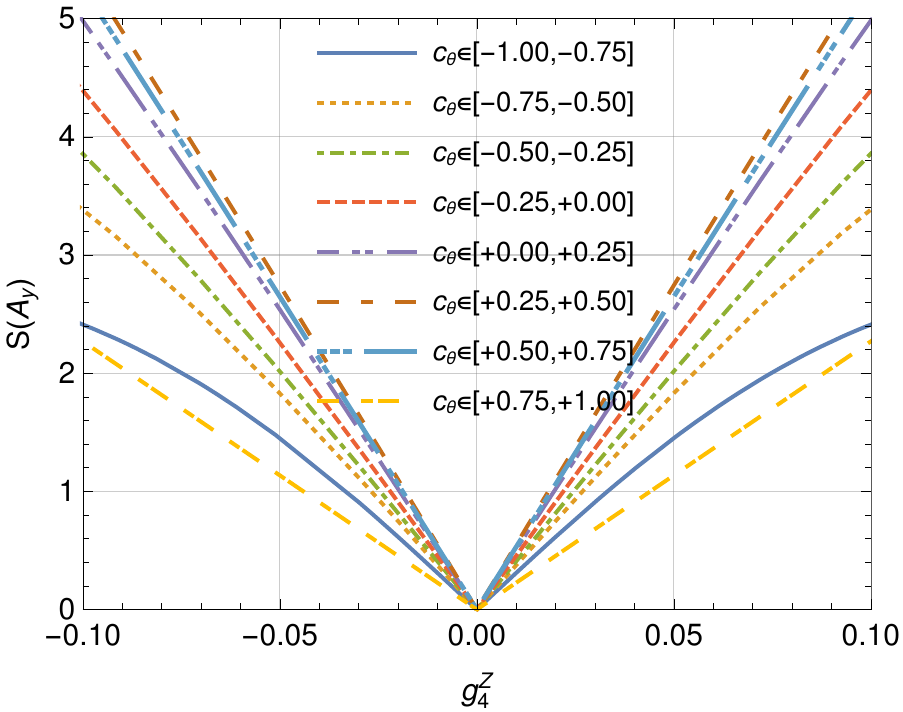}
	\includegraphics[width=0.49\textwidth]{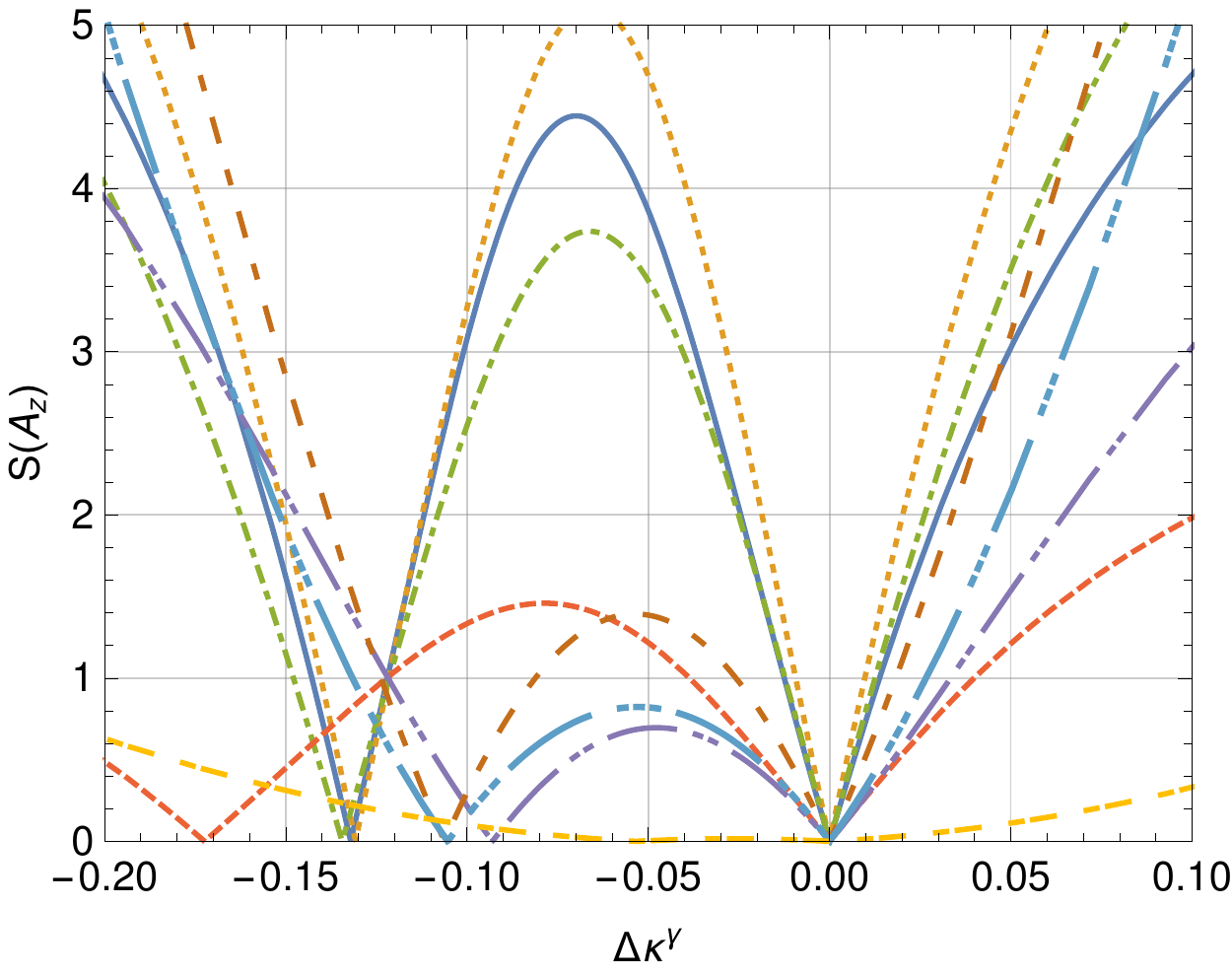}
	\caption{\label{fig:sensitivity_binned}The one-parameter  sensitivities of $A_x$ 
		on $g_4^Z$ ({\em left-panel}) and of $A_z$ on $\Delta\kappa^\gamma$ ({\em right-panel}) in eight bins
		at $\sqrt{s}=500$ GeV,  ${\cal L}=100$ fb$^{-1}$ with $c_\theta = \cos\theta_{W^-}$  for unpolarized beams.}
\end{figure*}
\begin{figure*}
	\centering
	\includegraphics[width=0.53\textwidth]{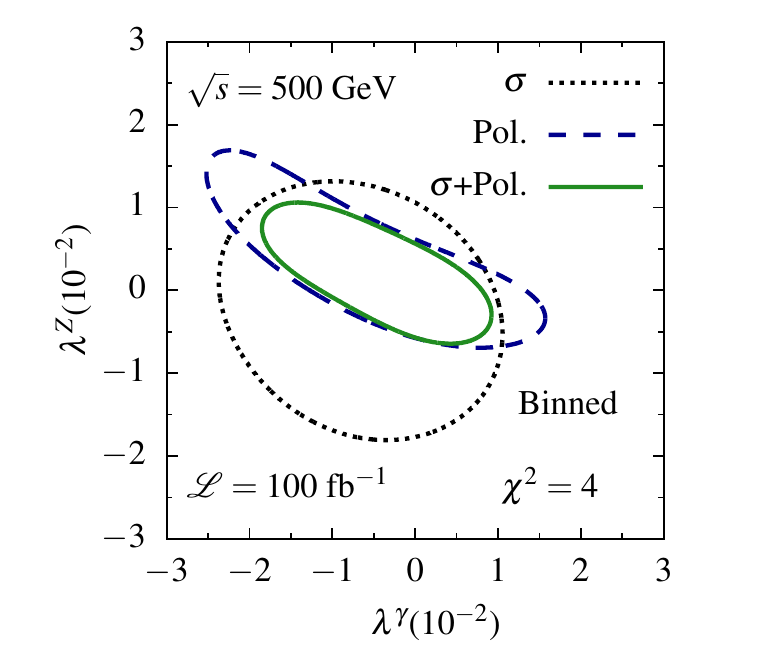}
	\includegraphics[width=0.45\textwidth]{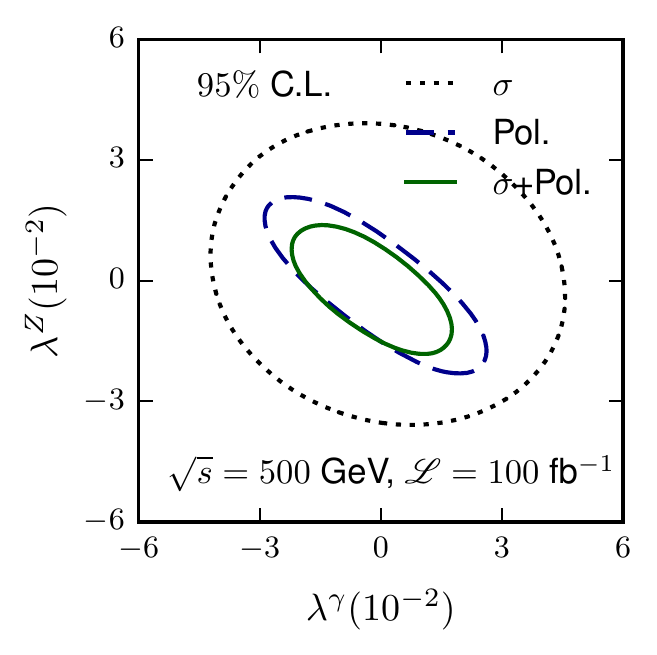}
	\caption{\label{fig:sig-vs-sigpol} The  $\chi^2=4$ contours in the {\em left-panel} and $95~\%$ C.L. contours from 
		simultaneous analysis in the  {\em right-panel} in the $\lambda^\gamma$--$\lambda^Z$ plane using the binned cross sections ($\sigma$)
		alone in {\it dotted} (black) lines, just binned polarizations asymmetries (Pol.) in {\it dashed} (blue) lines and the 
		binned cross sections together with binned polarization 
		asymmetries ($\sigma$ + Pol.) in {\it solid} (green) lines for $\sqrt{s}=500$ GeV, ${\cal L}=100$ fb$^{-1}$. } 
\end{figure*}
Here, we have a total of $14$ different anomalous couplings to be measured, while 
we only have $10$ observables. A certain combination of large couplings may 
mimic the SM within the statistical errors. To avoid these, we need more
 observables to be included in the analysis. We achieve this by dividing $\cos\theta_{W^-}$
 into eight bins and calculate the cross section and
polarization asymmetries in all of them.
In Fig.~\ref{fig:diffSigmaSM2}, the cross section and  the polarization asymmetries $A_z$, $A_{x}$, and $A_y$ are shown  as 
a function of $\cos\theta_{W^-}$   for the SM and some  {\tt aTGC} couplings
for both polarized and unpolarized beams.
The SM values for unpolarized cases are shown in {\it dotted} (blue) lines, and the
 SM values with a polarization of $(\eta_3,\xi_3)=(+0.6,-0.6)$ are shown in {\it dashed} (black) lines.
The {\it solid} (red) lines correspond to unpolarized {\tt aTGC} values, while   {\it dashed-dotted}
(green) lines represent polarized {\tt aTGC} values of observables. 
For the cross section ({\em left-top-panel}), we take $\Delta g_1^\gamma$  to be $0.1$  and all other 
couplings to zero  for both polarized and unpolarized beams.
We see that the fractional deviation from the SM value is larger in the
most backward bin ($\cos\theta_{W^-}\in(-1.0 , -0.75)$) and gradually reduces in the
forward direction.  The deviation is even larger in case
of beam polarization.
The sensitivity of the cross section on  $\Delta g_1^\gamma$  is thus expected to be high in the most 
backward bin. In the case of asymmetries, $A_z$ ({\em right-top-panel}), $A_{xz}$ ({\em left-bottom-panel}) 
and $A_y$ ({\em right-bottom-panel}), the {\tt aTGC} are assumed to be $\Delta\kappa^Z=0.05$, 
$\lambda^Z=0.05$  and $g_4^Z=0.05$, respectively, while others are kept at 
zero. The changes in the asymmetries due to {\tt aTGC} are larger in the backward
bin for both  polarized  and unpolarized beam cases. 
We note that the asymmetries may not have the highest sensitivity in the most backward bin but in
 some other bin. We consider 
the cross section and eight polarization asymmetries  in all eight bins, i.e., we have  $72$ observables in our
analysis.

one-parameter sensitivities of the set of nine observables in all  eight bins have
been studied. We show the sensitivity of $A_y$  on $g_4^Z$ and of $A_z$ on $\Delta\kappa^\gamma$ 
in the eight bins in Fig.~\ref{fig:sensitivity_binned} as representative. The tightest limits based on 
the sensitivity (coming from one bin)  is roughly twice as tight as  compared 
to the unbinned case in Fig.~\ref{fig:sensitivity}. Thus, we expect simultaneous limits
on all the couplings to be tighter when using binned observables.

\begin{table*}[!ht]\caption{\label{tab:obs-analysis-name} The list  of  
        analyses performed in the present work
        and  set of  observables used with  different kinematical cuts   to obtain 
        simultaneous limits on the anomalous couplings  
        at $\sqrt{s}=500$ GeV, ${\cal L}=100$ fb$^{-1}$ with unpolarized beams.
        The rectangular volumes of couplings at $95\%$ BCI are 
        shown in the last column for each analyses (see text
        for details).  }
    \renewcommand{\arraystretch}{1.50}
    \begin{tabular*}{\textwidth}{@{\extracolsep{\fill}}llll@{}}\hline
        Analysis Name  & Set of Observables & Kinematical Cut on $\cos\theta_{W^-}$  & Volume of Limits \\\hline
        {\tt $\sigma$-unbinned} & $\sigma$ & $\cos\theta_{W^-}\in[-1.0,1.0]$& $4.4\times 10^{-11}$\\
        {\tt Unbinned} & 
        $\sigma$, $A_{fb}$, $A_i$
        & $\cos\theta_{W^-}\in[-1.0,1.0]$ & $3.1\times 10^{-12}$\\
        {\tt $\sigma$-binned} &  $\sigma$ & 
        $\cos\theta_{W^-}\in[\frac{m-5}{4},\frac{m-4}{4}]$, $m=1,2,\dots, 8$
        & $3.7 \times 10^{-12}$ \\
 {\tt Pol.-binned} &$A_i$ &$\cos\theta_{W^-}\in[\frac{m-5}{4},\frac{m-4}{4}]$, $m=1,2,\dots, 8$ & $1.6\times 10^{-15} $ \\
        {\tt Binned} & $\sigma$,  $A_i$ &  $\cos\theta_{W^-}\in[\frac{m-5}{4},\frac{m-4}{4}]$, $m=1,2,\dots, 8$& $5.2 \times 10^{-17}$\\
        \hline
    \end{tabular*}
\end{table*}
We perform a set of Markov-chain--Monte-Carlo (MCMC) analyses with a different set of observables for different kinematical
cuts with unpolarized beams to understand their roles in providing limits on
the anomalous couplings. These analyses are listed in 
Table~\ref{tab:obs-analysis-name}. 
The corresponding $14$-dimensional rectangular 
volume\footnote[3]{This volume of limit  is  
     the volume of a $14$-dimensional rectangular box bounding by the 
    $95\%$ BCI projection of simultaneous limits in each coupling, 
    which can be a measure of goodness of the benchmark beam
    polarization. We computed the cross section and other asymmetries keeping 
    term up to quadratic in couplings. In this case, even a single observable 
    can  give a finite volume of limit and constrain all $14$ couplings, which 
    would not be possible if only terms linear in couplings were present.}  made out 
of $95\%$ Bayesian confidence
interval (BCI) on the anomalous couplings are also listed in
Table~\ref{tab:obs-analysis-name} in the last column. 
The simplest analysis would be to consider only the cross section in the full $\cos\theta_{W^-}$ domain
and perform MCMC analysis which is named as  {\tt $\sigma$-unbinned}. The typical
$95\%$ limits on the parameters range from $\sim \pm0.04$ to $\pm0.25$ giving the volume of 
limits to be $4.4\times 10^{-11}$. As we have polarizations asymmetries, the straight
forward analysis would be to consider all the observables  
for the full domain of $\cos\theta_{W^-}$. This analysis is named  {\tt Unbinned} where  limits on 
anomalous couplings get constrained better reducing the volume of limits by a factor of $10$
compared to the {\tt $\sigma$-unbinned}.  
To see how binning improves the limits, we perform an analysis named {\tt $\sigma$-binned} 
using only the cross section in eight bins. We see that the analysis {\tt $\sigma$-binned} is better 
than the {\tt $\sigma$-unbinned} and comparable to the analysis {\tt Unbinned}.
To see the strength of the polarization asymmetries, we perform an analysis named {\tt Pol.-binned} using
just the polarization asymmetries in eight bins. We see that this analysis is much better than the 
analysis {\tt $\sigma$-binned}.
The most natural and complete analysis would be to  consider all the  observables after binning.
The analysis is named as  {\tt Binned} which has  limits  much better than any analysis. 
The comparison between  the analyses, {\tt $\sigma$-binned}, {\tt Pol.-binned}, and {\tt Binned}
is shown in Fig.~\ref{fig:sig-vs-sigpol} in the panel $\lambda^\gamma$--$\lambda^Z$ in two-parameter
({\em left-panel}) as well as in multi-parameter ({\em right-panel}) analysis using MCMC as 
representative. The {\em right-panel} reflects  Table~\ref{tab:obs-analysis-name}. The behaviours are same
even in the two parameter analysis  ({\em left-panel}) by keeping all other parameter to zero, i.e, the bounded region for $\chi^2=4$ is smaller in   {\tt Pol.-binned} (Pol.) than  {\tt $\sigma$-binned} ($\sigma$)
and smallest for {\tt Binned} ($\sigma$+Pol.).

We also calculate one-parameter limits on all the couplings at $95~\%$ C.L.  
considering all the binned observables with 
unpolarized beams in the effective vertex formalism as well as in the effective operator 
approach and list them in the last column of  Tables~\ref{tab:Limits-Lag} \&~\ref{tab:Limits-Op}, respectively, for comparison.
In the next subsection, we study the effect of beam polarizations on the limits of the anomalous couplings. 
\subsection{Effect of beam polarizations to the limits on the anomalous couplings}\label{sec:3.2} 
\begin{table*}
 \caption{\label{tab:Limits-Lag}List of  posterior $95~\%$ BCI
        of anomalous couplings $c_i^{\cal L}$ ($10^{-2}$) of the Lagrangian in Eq.~(\ref{eq:Lagrangian})   at $\sqrt{s}=500$ GeV, ${\cal L}=100$ fb$^{-1}$  for a chosen set of longitudinal beam polarizations $\eta_3$ and $\xi_3$ 
        from MCMC global fits in {\tt Binned} case. The limits for the best choice of beam polarization within technological reach, i.e., $(\pm0.8,\mp0.6)$ are marked in \textbf{bold}. The pictorial visualization for these $95~\%$ BCI of $c_i^{\cal L}$   
        are shown in Fig.~\ref{fig:Limits-combined} in the {\em left-panel}. 
        The one-parameter ($1P$) limits ($10^{-2}$)   at $95~\%$ BCI  with unpolarized beams are given in the
        last column for comparison. The notation used here is $_{ low}^{ high}\equiv [ low,  high]$ with
        $low$ being the lower limit and $high$ being the upper limit.}
    \renewcommand{\arraystretch}{1.50}
  \centering
\begin{tabular*}{\textwidth}{@{\extracolsep{\fill}}cccccccc@{}}\hline
Parameter             &$(0,0)          $&$(\pm 0.2,\mp 0.2)  $&$(\pm 0.4,\mp 0.4)  $&$(\pm 0.6,\mp 0.6)   $&          $ \mathbf{ (\pm 0.8,\mp 0.6)} $&$(\pm 0.8,\mp 0.8) $&$1P$$( 0, 0)   $\\ \hline
$ \Delta g_1^{\gamma}         $&$ _{ -8.5 }^{+5.5 }$&$ _{-7.4 }^{+3.3 }$&$ _{-6.0 }^{+ 2.7 }$&$ _{ -2.7 }^{+2.1 }$&$ \mathbf{_{-2.3 }^{+ 1.7 }} $&$ _{ -2.0 }^{+1.6 }$&$ _{-1.4 }^{+1.3 }$ \\  \hline 
$ g_4^{\gamma}                $&$ _{ -6.0 }^{+5.8 }$&$ _{-5.4 }^{+5.3 }$&$ _{-4.0 }^{+ 4.0 }$&$ _{ -3.0 }^{+3.0 }$&$ \mathbf{_{-2.5 }^{+ 2.5 }} $&$ _{ -2.2 }^{+2.2 }$&$ _{-1.9 }^{+1.9 }$ \\  \hline 
$ g_5^{\gamma}                $&$ _{ -6.1 }^{+6.1 }$&$ _{-5.2 }^{+5.1 }$&$ _{-3.1 }^{+ 2.6 }$&$ _{ -2.0 }^{+1.4 }$&$ \mathbf{_{-1.6 }^{+ 1.1 }} $&$ _{ -1.4 }^{+1.0 }$&$ _{-2.0 }^{+1.9 }$ \\  \hline 
$ \lambda^{\gamma}           $&$ _{ -1.8 }^{+1.4 }$&$ _{-1.6 }^{+1.2 }$&$ _{-1.2 }^{+ 1.2 }$&$ _{ -0.68}^{+1.0}$&$ \mathbf{_{-0.61}^{+ 0.89}} $&$ _{ -0.57}^{+0.81}$&$ _{-1.1 }^{+0.77}$ \\  \hline 
$ \widetilde{\lambda^{\gamma}}$&$ _{ -1.6 }^{+1.6 }$&$ _{-1.4 }^{+1.4 }$&$ _{-1.1 }^{+ 1.1 }$&$ _{ -0.88}^{+0.88}$&$ \mathbf{_{-0.82}^{+ 0.82}} $&$ _{ -0.78}^{+0.77}$&$ _{-1.0 }^{+1.0 }$ \\  \hline 
$  \Delta\kappa^{\gamma}      $&$ _{ -5.7 }^{+0.91}$&$ _{-4.4 }^{+0.32}$&$ _{-4.3 }^{+ 0.46}$&$ _{ -0.69}^{+0.28}$&$ \mathbf{_{-0.55}^{+ 0.27}} $&$ _{ -0.48}^{+0.25}$&$ _{-0.34}^{+0.33}$ \\  \hline 
$ \widetilde{\kappa^{\gamma}} $&$ _{ -6.0 }^{+6.1 }$&$ _{-5.2 }^{+5.2 }$&$ _{-3.9 }^{+ 4.0 }$&$ _{ -3.0 }^{+2.9 }$&$ \mathbf{_{-2.6 }^{+ 2.6 }} $&$ _{ -2.3 }^{+2.3 }$&$ _{-2.4 }^{+2.3 }$ \\  \hline 
$ \Delta g_1^Z                $&$ _{ -3.7 }^{+7.2 }$&$ _{-2.8 }^{+5.6 }$&$ _{-2.6 }^{+ 4.5 }$&$ _{ -2.0 }^{+2.1 }$&$ \mathbf{_{-1.7 }^{+ 1.8 }} $&$ _{ -1.5 }^{+1.6 }$&$ _{-1.3 }^{+1.3 }$ \\  \hline 
$ g_4^Z                       $&$ _{ -4.7 }^{+4.8 }$&$ _{-4.3 }^{+4.3 }$&$ _{-3.3 }^{+ 3.3 }$&$ _{ -2.5 }^{+2.5 }$&$ \mathbf{_{-2.2 }^{+ 2.2 }} $&$ _{ -2.0 }^{+2.0 }$&$ _{-1.4 }^{+1.4 }$ \\  \hline 
$ g_5^Z                       $&$ _{ -4.8 }^{+4.7 }$&$ _{-4.1 }^{+4.0 }$&$ _{-2.3 }^{+ 2.1 }$&$ _{ -1.5 }^{+1.3 }$&$ \mathbf{_{-1.3 }^{+ 1.0 }} $&$ _{ -1.2 }^{+0.86}$&$ _{-1.3 }^{+1.2 }$ \\  \hline 
$ \lambda^Z                   $&$ _{ -1.5 }^{+1.1 }$&$ _{-1.3 }^{+1.0 }$&$ _{-1.1 }^{+ 0.80}$&$ _{ -0.94}^{+0.49}$&$ \mathbf{_{-0.83}^{+ 0.47}} $&$ _{ -0.76}^{+0.44}$&$ _{-0.57}^{+0.56}$ \\  \hline 
$ \widetilde{\lambda^Z}       $&$ _{ -1.3 }^{+1.3 }$&$ _{-1.1 }^{+1.1 }$&$ _{-0.90}^{+ 0.90}$&$ _{ -0.77}^{+0.77}$&$ \mathbf{_{-0.73}^{+ 0.73}} $&$ _{ -0.68}^{+0.68}$&$ _{-0.56}^{+0.57}$ \\  \hline 
$ \Delta\kappa^Z              $&$ _{ -1.5 }^{+3.6 }$&$ _{-0.49}^{+3.2 }$&$ _{-0.44}^{+ 3.1 }$&$ _{ -0.38}^{+0.56}$&$ \mathbf{_{-0.35}^{+ 0.43}} $&$ _{ -0.32}^{+0.36}$&$ _{-0.48}^{+0.43}$ \\  \hline 
$ \widetilde{\kappa^Z}        $&$ _{ -5.0 }^{+4.7 }$&$ _{-4.2 }^{+4.2 }$&$ _{-3.3 }^{+ 3.3 }$&$ _{ -2.5 }^{+2.5 }$&$ \mathbf{_{-2.2 }^{+ 2.2 }} $&$ _{ -2.0 }^{+2.1 }$&$ _{-1.5 }^{+1.5 }$ \\  \hline
\end{tabular*}
 \end{table*}
\begin{table*}
\caption{\label{tab:Limits-Op} The list of  posterior   $95~\%$ BCI
of anomalous couplings $c_i^{\cal O}$ (TeV$^{-2}$)  of effective operators  in Eq.~(\ref{eq:opertaors-dim6}) and their translated limits
on the couplings $c_i^{{\cal L}_g}$ ($10^{-2}$)
for $\sqrt{s}=500$ GeV, ${\cal L}=100$ fb$^{-1}$ in the 
{\tt Binned} case for a chosen set of longitudinal beam polarizations $\eta_3$ and $\xi_3$ from MCMC global fits.
The pictorial visualization for these $95~\%$ BCI of $c_i^{\cal O}$ and $c_i^{{\cal L}_g}$ are shown in   Fig.~\ref{fig:Limits-combined} in the 
{\em right-top} and {\em right-bottom} panels, respectively. The rest of the   details are same 
as in Table~\ref{tab:Limits-Lag}. }
\renewcommand{\arraystretch}{1.50}
\begin{tabular*}{\textwidth}{@{\extracolsep{\fill}}cccccccc@{}}\hline
Parameter               &$(0,0)          $&$(\pm 0.2,\mp 0.2)  $&$(\pm 0.4,\mp 0.4)  $&$(\pm 0.6,\mp 0.6)   $                    &$\mathbf{ (\pm 0.8,\mp 0.6)  }$&$(\pm 0.8,\mp 0.8) $&$1P$$( 0, 0)   $\\ \hline
$\frac{c_{WWW}}{\Lambda^2}            $&$ _{ -1.9 }^{+1.3 }$&$ _{ -1.4 }^{+ 1.2 }$&$ _{ -1.1 }^{+1.2 }$&$ _{-0.96 }^{+1.1  }$&$\mathbf{ _{ -1.0  }^{+1.1  }}$&$ _{ -0.94 }^{+ 1.0  }$&$ _{-0.97}^{+ 0.84 }$ \\ \hline
$\frac{c_{W}}{\Lambda^2}              $&$ _{ -1.4 }^{+5.0 }$&$ _{ -1.1 }^{+ 4.6 }$&$ _{ -0.86}^{+0.83}$&$ _{-0.72 }^{+0.58 }$&$\mathbf{ _{ -0.73 }^{+0.60 }}$&$ _{ -0.63 }^{+ 0.55 }$&$ _{-0.58}^{+ 0.55 }$ \\ \hline
$\frac{c_{B}}{\Lambda^2}              $&$ _{-23.7 }^{+2.7 }$&$ _{-20.2 }^{+ 1.9 }$&$ _{ -1.3 }^{+0.98}$&$ _{-0.75 }^{+0.62 }$&$\mathbf{ _{ -0.64 }^{+0.56 }}$&$ _{ -0.53 }^{+ 0.47 }$&$ _{-1.3 }^{+ 1.2 }$ \\ \hline
$\frac{c_{\widetilde{WWW}}}{\Lambda^2}$&$ _{ -1.4 }^{+1.4 }$&$ _{ -1.1 }^{+ 1.1 }$&$ _{ -0.97}^{+0.97}$&$ _{-0.93 }^{+0.94 }$&$\mathbf{ _{ -0.90 }^{+0.91 }}$&$ _{ -0.87 }^{+ 0.87 }$&$ _{-0.98}^{+ 0.97 }$ \\ \hline
$\frac{c_{\widetilde{W}}}{\Lambda^2}  $&$ _{-12.0 }^{+2.1 }$&$ _{-10.0 }^{+ 9.8 }$&$ _{ -6.7 }^{+6.6 }$&$ _{-4.1  }^{+4.2  }$&$\mathbf{ _{ -3.2  }^{+3.2  }}$&$ _{ -2.6  }^{+ 2.6  }$&$ _{-9.9 }^{+ 10.1 }$ \\ \hline  \hline  
$\lambda^{V}                 $&$ _{-0.79 }^{+ 0.52 }$&$ _{-0.58 }^{+0.50 }$&$ _{-0.46 }^{+ 0.49 }$&$ _{-0.40 }^{+0.46 }$&$ \mathbf{_{-0.41 }^{+ 0.45 }}$&$ _{ -0.39 }^{+0.42 }$&$ _{-0.40 }^{+0.35 }$\\  \hline 
$\widetilde{\lambda^{V}}     $&$ _{-0.60 }^{+ 0.60 }$&$ _{-0.45 }^{+0.44 }$&$ _{-0.40 }^{+ 0.40 }$&$ _{-0.38 }^{+0.39 }$&$ \mathbf{_{-0.37 }^{+ 0.37 }}$&$ _{ -0.36 }^{+0.36 }$&$ _{-0.41 }^{+0.40 }$\\  \hline 
$\Delta\kappa^{\gamma}       $&$ _{-6.4  }^{+ 0.52 }$&$ _{-5.1  }^{+0.44 }$&$ _{-0.38 }^{+ 0.28 }$&$ _{-0.32 }^{+0.24 }$&$ \mathbf{_{-0.32 }^{+ 0.25 }}$&$ _{ -0.28 }^{+0.23 }$&$ _{-0.61 }^{+0.56 }$\\  \hline 
$\widetilde{\kappa^{\gamma}} $&$ _{-3.9  }^{+ 3.9  }$&$ _{-3.2  }^{+3.2  }$&$ _{-2.1  }^{+ 2.1  }$&$ _{-1.3  }^{+1.3  }$&$ \mathbf{_{-1.0  }^{+ 1.0  }}$&$ _{ -0.84 }^{+0.84 }$&$ _{-3.2}^{+3.2}$\\  \hline 
$\Delta g_1^Z                $&$ _{-0.59 }^{+ 2.1  }$&$ _{-0.45 }^{+1.9  }$&$ _{-0.36 }^{+ 0.34 }$&$ _{-0.30 }^{+0.24 }$&$ \mathbf{_{-0.30 }^{+ 0.25 }}$&$ _{ -0.26 }^{+0.23 }$&$ _{-0.24 }^{+0.23 }$\\  \hline 
$\Delta\kappa^Z              $&$ _{-0.73 }^{+ 3.6  }$&$ _{-0.45 }^{+3.2  }$&$ _{-0.33 }^{+ 0.34 }$&$ _{-0.24 }^{+0.21 }$&$ \mathbf{_{-0.24 }^{+ 0.21 }}$&$ _{ -0.20 }^{+0.19 }$&$ _{-0.30 }^{+0.30 }$\\  \hline 
$\widetilde{\kappa^Z}        $&$ _{-1.1  }^{+ 1.1  }$&$ _{-0.91 }^{+0.92 }$&$ _{-0.61 }^{+ 0.62 }$&$ _{-0.38 }^{+0.38 }$&$ \mathbf{_{-0.30 }^{+ 0.29 }}$&$ _{ -0.24 }^{+0.24 }$&$ _{-0.93 }^{+0.92 }$\\  \hline 
\end{tabular*}                                                                                                             
 \end{table*}                                                                                                       
\begin{figure*}
    \begin{minipage}{0.496\textwidth}
    \centering
        \includegraphics[width=\textwidth]{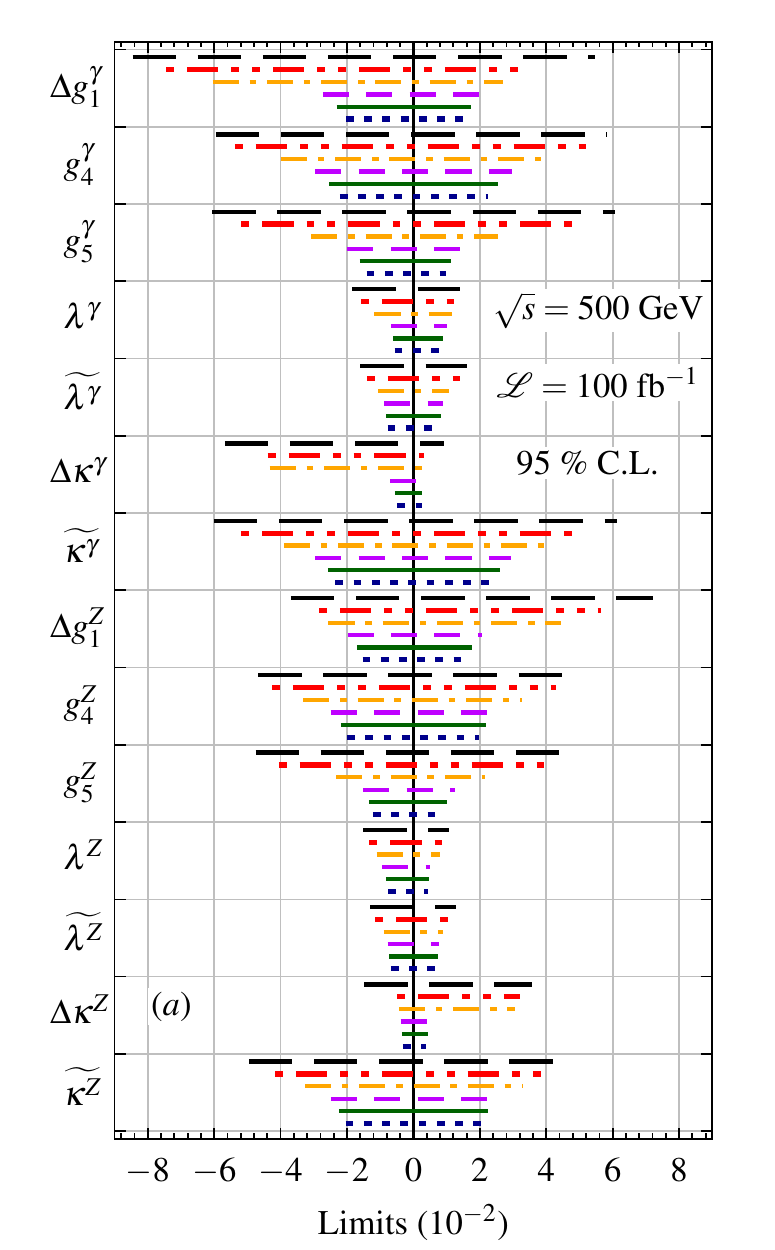}
    \end{minipage}
    \begin{minipage}{0.496\textwidth}
    \centering
        \includegraphics[width=\textwidth]{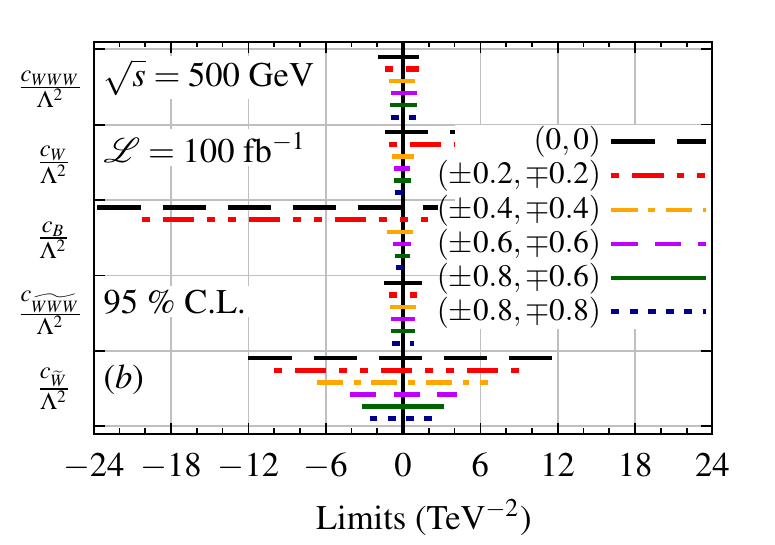}
        \includegraphics[width=\textwidth]{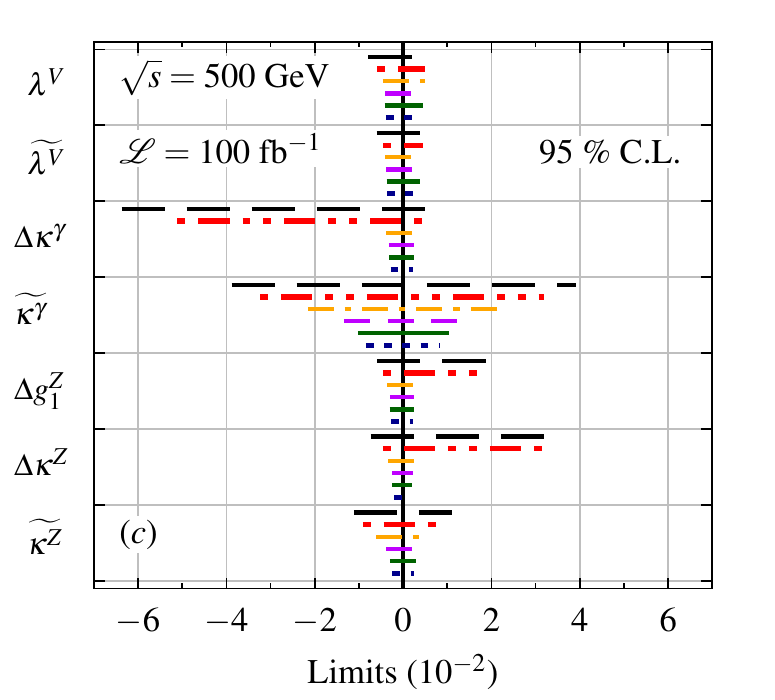}
    \end{minipage}
    \caption{\label{fig:Limits-combined} The pictorial visualizations of $95~\%$ BCI limits obtained from from MCMC global fits  $(a):$ on the  anomalous couplings $c_i^{\cal L}$ 
        in the {\em left-panel}, $(b):$ on  $c_i^{\cal O}$ in the {\em right-top-panel} and $(c):$ on $c_i^{{\cal L}_g}$ in the {\em right-bottom-panel}
         for $\sqrt{s}=500$ GeV, ${\cal L}=100$ 
        fb$^{-1}$ using the binned observables. The numerical values of the 
        limits  can be read of  in Tables~\ref{tab:Limits-Lag} \&~\ref{tab:Limits-Op}.} 
\end{figure*}
\begin{figure*}
    \centering
    \includegraphics[width=0.9\textwidth]{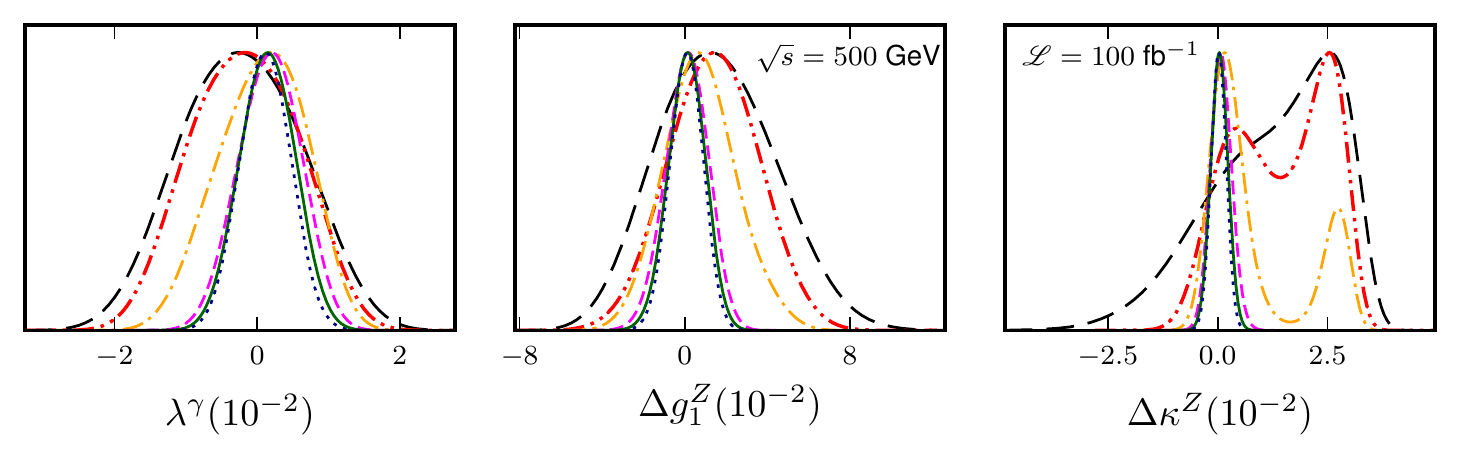}
    \includegraphics[width=0.325\textwidth]{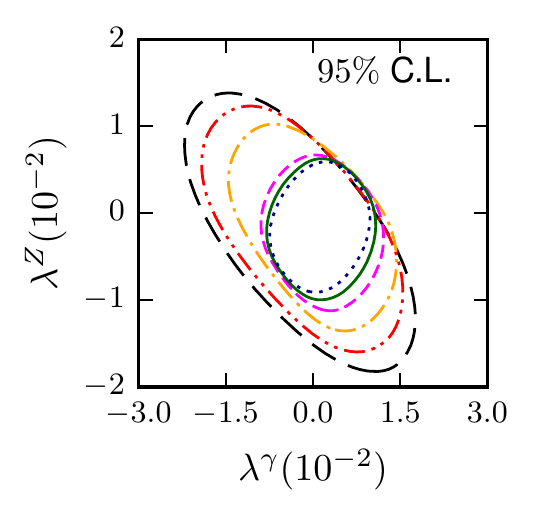}
    \includegraphics[width=0.325\textwidth]{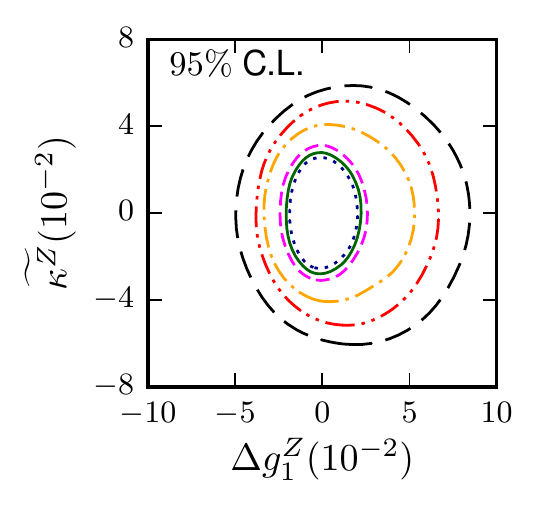}
    \includegraphics[width=0.325\textwidth]{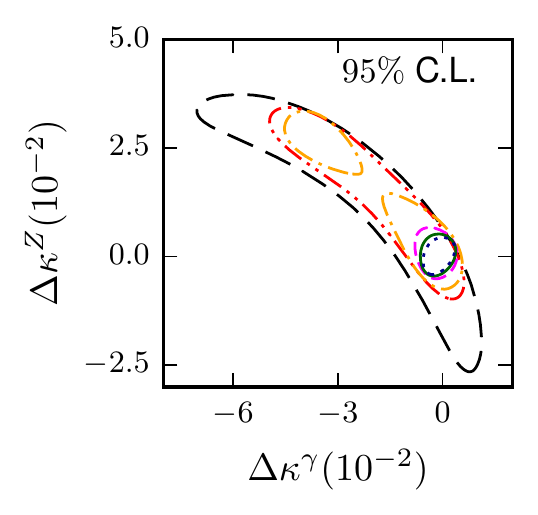}
    \caption{\label{fig:MCMC-beamPol-Lag} The marginalized $1D$ projections for the couplings
        $\lambda^\gamma$, $\Delta g_1^Z$ and $\Delta\kappa^Z$ in the {\em top-panel} 
        and $2D$ projections at $95~\%$ C.L. on 
        $\lambda^\gamma$--$\lambda^Z$, $\Delta g_1^Z$--$\wtil{\kappa^Z}$  and   
        $\Delta\kappa^\gamma$--$\Delta\kappa^Z$ planes in the {\em bottom-panel}     
        from MCMC for a set of choice of beam polarizations are shown
    for $\sqrt{s}=500$ GeV, ${\cal L}=100$ fb$^{-1}$ using the binned observables in 
    the effective vertex formalism. The legend labels are same as in Figs.~\ref{fig:Limits-combined} \&~\ref{fig:MCMC-BeamPol-Op}. } 
\end{figure*}
\begin{figure*}
\centering
    \includegraphics[width=0.96\textwidth]{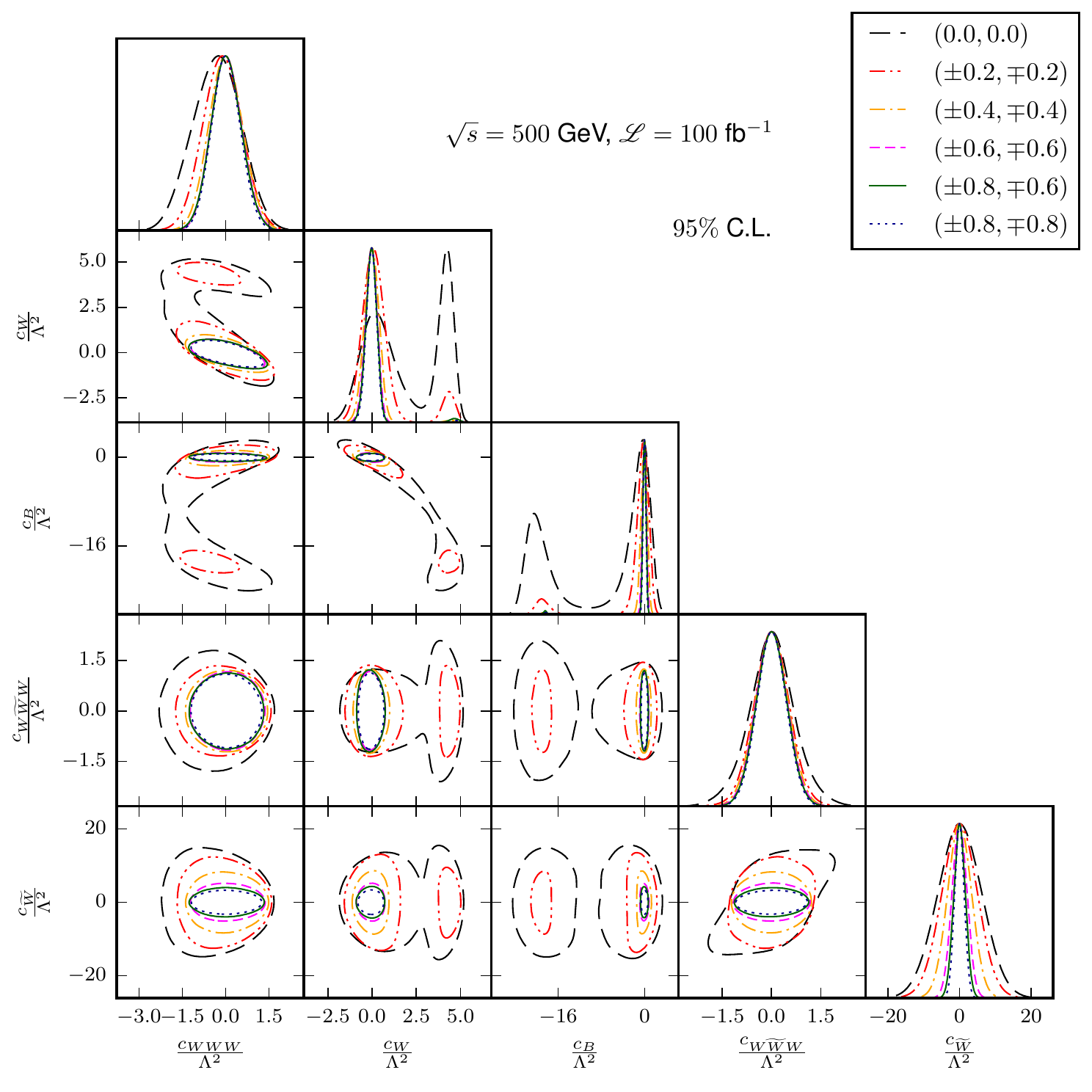}
    \caption{\label{fig:MCMC-BeamPol-Op} All the  marginalized $1D$ projections and $2D$ projections at $95~\%$ C.L. from the MCMC in a triangular array
    for the effective operators (TeV$^{-2}$) for a set of choices of beam polarizations
    for $\sqrt{s}=500$ GeV, ${\cal L}=100$ fb$^{-1}$ using the binned observables. } 
\end{figure*}

We perform a MCMC analysis to estimate simultaneous limits on the anomalous couplings using the binned observables in both
effective vertex formalism with $14$ independent couplings and an effective operator approach
with five independent couplings for a set of chosen beam polarizations $(\eta_3,\xi_3)$ to be
$(0,0)$, $(+0.2,-0.2)$, $(+0.4,-0.4)$, $(+0.6,-0.6)$, $(+0.8,-0.6)$, $(+0.8,-0.8)$ along with their
opposite values. The beam polarization $(+\eta_3,+\xi_3)$ and its opposite $(-\eta_3,-\xi_3)$
are combined at the level of $\chi^2$ as 
\begin{eqnarray}\label{eq:beampol-combine}
\chi^2_{tot}(\pm\eta_3,\pm\xi_3)= \sum_{bin}  \sum_{N} \left(\chi^2\left[{\cal O}_N(+\eta_3,+\xi_3)\right] \right.\nonumber \\
\left.+\chi^2\left[{\cal O}_N(-\eta_3,-\xi_3)\right] \right) ,
\end{eqnarray}
where $N$ runs over all the observables. The $95~\%$ BCI simultaneous limits for the chosen set of 
beam polarizations combined according to Eq.~(\ref{eq:beampol-combine}) are shown in 
Table~\ref{tab:Limits-Lag} for effective vertex formalism ($c_i^{\cal L}$) 
and in Table~\ref{tab:Limits-Op} for  effective operator approach ($c_i^{\cal O}$).  
The corresponding translated limit to the vertex factor couplings  $c_i^{{\cal L}_g}$ 
are also shown  in the  Table~\ref{tab:Limits-Op} using relation from 
Eq.~(\ref{eq:Operator-to-Lagrangian}). While presenting limits the  following notation is used  
$$_{ low}^{ high}\equiv [ low,  high]$$ with $low$ being lower limit and $high$ being upper limit.
A pictorial visualization of the limits shown in Table~\ref{tab:Limits-Lag} \& and \ref{tab:Limits-Op}
is given in Fig.~\ref{fig:Limits-combined} for the easy comparisons. 
The limits on the couplings get tighter as the magnitude of the beam polarizations are increased along $\eta_3=-\xi_3$ path
and become tightest at the extreme beam polarization 
$(\pm 0.8, \mp 0.8)$. However, the choice $(\pm 0.8, \mp 0.6)$ is best to put constraints on the couplings
 within the technological reach~\cite{Vauth:2016pgg,MoortgatPick:2006qp}. 

To show the effect of beam polarizations the marginalized $1D$ projection for the couplings
$\lambda^\gamma$, $\Delta g_1^Z$ and $\Delta\kappa^Z$ as well as $2D$ projection at $95~\%$ C.L. on 
$\lambda^\gamma$--$\lambda^Z$, $\Delta g_1^Z$--$\wtil{\kappa^Z}$ and   
$\Delta\kappa^\gamma$--$\Delta\kappa^Z$  planes are shown in Fig.~\ref{fig:MCMC-beamPol-Lag} for the 
effective vertex formalism ($c_i^{\cal L}$) as representative. We observe that as the magnitude of beam
polarizations are increased from $(0,0)$ to $(\pm 0.8, \mp 0.8)$ the contours get smaller centerd 
around the SM values in the $2D$ projection which is reflected in the $1D$ projection as well.
In the $\Delta\kappa^\gamma$--$\Delta\kappa^Z$ panel, the contours get divided into two part 
at $(\pm 0.4, \mp 0.4)$ and become one single contour later  centerd around the SM values. 
In the case of effective operator approach ($c_i^{\cal O} $), all the $1D$ and  $2D$ ($95~\%$ C.L.)  projections 
after marginalization are shown in Fig.~\ref{fig:MCMC-BeamPol-Op}. In this case the couplings
$c_W$ and $c_B$ has two patches up-to beam polarization $(\pm 0.2, \mp 0.2)$  and become one single
patch starting at beam polarization $(\pm 0.3, \mp 0.3)$ centerd around the SM values. As the magnitude of beam
polarizations are increased along the $\eta_3=-\xi_3$ line, the measurement of the anomalous couplings
gets improved. The set of beam polarizations chosen here are mostly  along $\eta_3=-\xi_3$ line, but some
choices off to the line might provide the same results. A discussion on the choice of beam polarization is given in the next subsection.
\subsection{On the choice of beam polarizations}\label{sec:3.3}
\begin{figure*}
    \centering
    \includegraphics[width=0.6\textwidth]{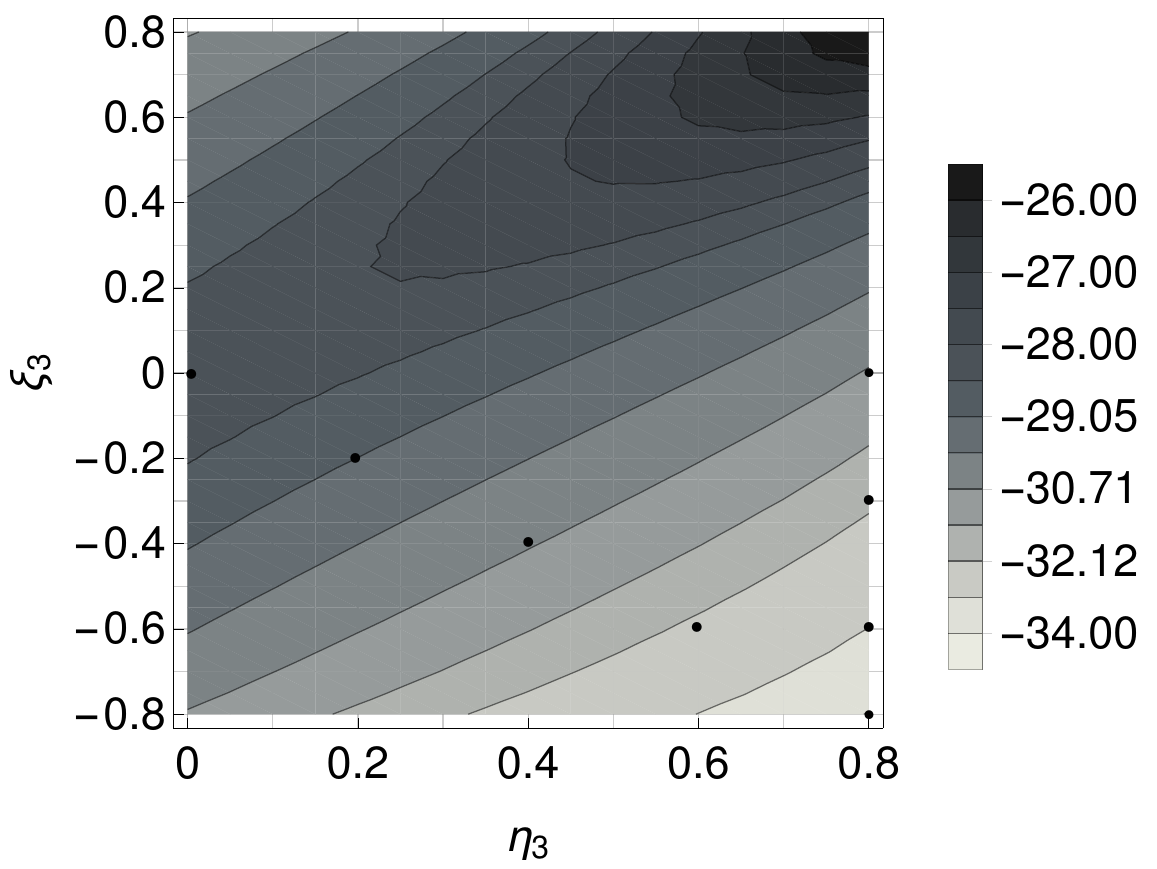}
    \caption{\label{fig:averageLikelihood}The  averaged likelihood 
        $L_{Av}=L(V_{\vec{f}}~;\eta_3,\xi_3)$ 
        in log scale as a function of $(\pm\eta_3 ,\pm\xi_3)$ in the
         effective vertex formalism for $\sqrt{s}=500$ GeV, ${\cal L}=100$ fb$^{-1}$.}
\end{figure*}
In the previous subsection, we found that $(\pm\eta_3 ,\pm\xi_3)=(\pm 0.8, \pm 0.6)$ is the best choice of beam
polarizations to provide simultaneous limits on the anomalous couplings obtained by MCMC analysis.
Here, we discuss the average likelihood or the
weighted volume of the parameter space defined as~\cite{Rahaman:2017qql}
\begin{eqnarray}\label{eq:average_likelihood}
L(V_{\vec{f}}~;\eta_3,\xi_3)&=&\int_{V_{\vec{f}}}  \exp{\left[-\frac{1}{2}\chi_{tot}^2(\vec{f},\eta_3,\xi_3)\right]} d\vec{f}
\end{eqnarray}
to cross-examine the beam polarization choices made in the previous section. 
Here $\vec{f}$ is the coupling vector and $V_{\vec{f}}$ is the volume 
of parameter space over which the average is done and
$L(V_{\vec{f}}~;\eta_3,\xi_3)$ corresponds to  
 the volume of the parameter space that is statistically consistent with the SM . One naively expects the limits to be tightest when
$L(V_{\vec{f}}~;\eta_3,\xi_3)$ is minimum.
We calculate the above quantity as a function of $(\pm\eta_3,\pm\xi_3)$ 
for the   {\tt Binned} case in the
effective vertex formalism given in the  Lagrangian in Eq.~(\ref{eq:Lagrangian}) and  present it in Fig.~\ref{fig:averageLikelihood}.
As the opposite beam polarizations are combined, only the half-portions
are shown in the $\eta_3$--$\xi_3$ plane. The dot ({\tiny$\bullet$}) points along the $\eta_3=-\xi_3$ are the 
chosen choice of beam polarizations for the MCMC analysis.
We see that the average likelihood decreases along the $\eta_3=-\xi_3$ line while it increases
along the  $\eta_3=\xi_3$ line. The constant lines or contours of average likelihood in the figure imply that any beam polarizations
along the lines/contours will provide the similar shape of $1D$ and $2D$ projections of couplings and  their limits. For example, 
the point $(\pm 0.8,\mp 0.6)$ is equivalent to the point $(\pm 0.7,\mp 0.7)$ as well as 
$(\pm 0.6,\pm 0.8)$ roughly in providing simultaneous limits which are verified from the limits obtained by the  MCMC analysis.
From the figure, it is clear that the polarization $(\pm 0.8,\mp 0.6)$ is indeed the best choice to provide simultaneous limits on the anomalous couplings within the achievable range.
However, the plan for polarization choices are $(\eta_3,\xi_3)=(0,0)$, $(\pm 0.8,  0)$, $(\pm 0.8, \mp 0.3)$, and $(\pm 0.8, \mp 0.6)$ at the ILC~\cite{Barklow:2015tja,Bambade:2019fyw}.  These off-diagonal choices are equivalent to the diagonal choices we have used as  Fig.~\ref{fig:averageLikelihood} indicates. 
The polarization choice $(\pm 0.8,  0)$ is equivalent to  $(\pm 0.4, \mp 0.4)$  in providing limits on the couplings, while  
 $(\pm 0.8, \mp 0.3)$ is  equivalent to   $(\pm 0.6, \mp 0.6)$ ($(\pm 0.57, \mp 0.57)$ to be precise).  For completeness we also show the limits on the couplings  for the off-diagonal polarization choices $(\pm 0.8,  0)$ and $(\pm 0.8, \mp 0.3)$ in Table~\ref{tab:Limits-review-OpLag} on column $3$ and $5$, respectively in appendix~\ref{apendix:a}  in the $SU(2)\times U(1)$ gauge for ${\cal L}=100$ fb$^{-1}$. By comparing Tables~\ref{tab:Limits-Op} and~\ref{tab:Limits-review-OpLag}, one can confirm that the polarization choices $(\pm 0.8,  0)$ and $(\pm 0.8, \mp 0.3)$ are  indeed equivalent to the choices  $(\pm 0.4, \mp 0.4)$ and $(\pm 0.6, \mp 0.6)$, respectively. 
We also obtain limits on the couplings in the $SU(2)\times U(1)$ gauge  for the projected plan of the ILC~\cite{Barklow:2015tja}:  
polarization $(0,0)$ and $(\pm 0.8,0)$ at ${\cal L}=4$ ab$^{-1}$,  polarization  $(\pm 0.8,\mp 0.3)$ and  $(\pm 0.8,\mp 0.6)$ at ${\cal L}=3.2$ ab$^{-1}$ and show them in Table~\ref{tab:Limits-review-OpLag}. 
Increasing the luminosity from $100$ fb$^{-1}$ to the projected luminosity $3.2/4$ ab$^{-1}$ the limits on the couplings do not increase proportionately to the luminosity due to the systematic error considered here. If the systematic error is improved, we expect better limits on the couplings;  e.g.,  with no systematic error, the limits can be further  improved by a factor of $4$ at the projected luminosity. 


\section{Conclusion}\label{sec:conclusion}
In conclusion, we studied anomalous triple gauge boson couplings in 
$e^+e^-\to  W^+W^-$ with longitudinally polarized beams using    $W$ boson polarization observables
together with the total cross section and the forward-backward asymmetry
for $\sqrt{s}=500$ GeV and  luminosity of ${\cal L}=100$ fb$^{-1}$.
We have $14$ anomalous couplings, whereas we have 
only $10$ observables to measure them. So 
we binned all the observables ($A_{fb}$ excluded) in eight regions of the 
$\cos\theta_{W^-}$  to increase the number of observables to measure
the couplings. We estimated the simultaneous limit on all the couplings 
 for several chosen sets of beam polarization in both the effective vertex formalism and
 effective operator approach. The limits on the couplings are tighter when  $SU(2)\times U(1)$ symmetry is assumed. 
We show the consistency between
the best choice of beam polarizations and minimum likelihood averaged over the anomalous
couplings. We find  the polarization $(\pm 0.8,\mp 0.6)$ to be the best to provide the tightest
constraint on the anomalous couplings  at the ILC within the technological reach for both $100$ fb$^{-1}$ and $3.2$ ab$^{-1}$ of luminosity.
Our one-parameter limits  with unpolarized beams and simultaneous limits
for the best polarization choice at $100$  fb$^{-1}$ are already much better
than the one-parameter limits from experiments; see Table~\ref{tab:Limits-Op}.
Our analysis considers  certain simplifying assumptions,
such as the absence of initial-state/final-state radiation and detector effects. 
While the former might dilute the limits by  a small amount, the latter  is expected to have no effects on the results
as only the leptonic channel is assumed and no falvor tagging or reconstruction is required.

\vspace{0.5 cm}
\noindent \textbf{Acknowledgements} 
The authors thank Prof. Kaoru Hagiwara for useful discussions.
R.R. thanks the Department of Science 
and Technology, Government of India for support through the DST-INSPIRE Fellowship 
for doctoral program, INSPIRE CODE IF140075, 2014.  R.K.S. acknowledges SERB, DST, Government of India through the project EMR/2017/002778.
\appendix


\section{The dependences of observables  on anomalous  couplings $c_i^{\cal L}$ and limits on the couplings $c_i^{\cal O}/c_i^{{\cal L}_g}$ to the projected plan of the ILC}\label{apendix:a}
The anomalous gauge boson couplings $c_i^{\cal O}$ of the effective operator  
in Eq.~(\ref{eq:opertaors-dim6}),  the couplings $c_i^{\cal L}$ of the  
Lagrangian in Eq.~(\ref{eq:Lagrangian}), 
and the couplings $c_i^{{\cal L}_g}$ of the Lagrangian in the $SU(2)\times U(1)$ gauge (given in 
Eq.~(\ref{eq:Operator-to-Lagrangian}))  are labelled as
\begin{eqnarray}
c_i^{\cal O}&=&\{ c_{WWW}, c_{W}, c_B, c_{\wtil{WWW}}, c_{\wtil{W}} \}\label{eq:ciO} ,\\
c_i^{\cal L}&=&\{ \Delta g_1^V,g_4^V,g_5^V,\lambda^V,\wtil{\lambda^V},\Delta\kappa^V, \wtil{\kappa^V} \},~~~ V=\gamma,Z \label{eq:ciL},\\
c_i^{{\cal L}_g}&=& \{ \lambda^V, \wtil{\lambda^V}, 
\Delta\kappa^\gamma, \wtil{\kappa^\gamma}, \Delta g_1^Z,  \Delta\kappa^Z, \wtil{\kappa^Z} \}\label{eq:ciLg} .
\end{eqnarray}
The dependences of the observables on the anomalous couplings $c_i^{\cal L}$ are given in Table~\ref{tab:param_dependence}. The limits on the couplings $c_i^{\cal O}$ and $c_i^{{\cal L}_g}$ to the projected plan of the ILC are given in Table~\ref{tab:Limits-review-OpLag}.

\begin{table*}[h]
\caption{\label{tab:param_dependence} The dependence of observables 
        (numerators)  on the anomalous  couplings in the form of  $c_i^{\cal L}$ (linear), 
        $(c_i^{\cal L})^2$ (quadratic) and $c_i^{\cal L}c_j^{\cal L},~i\ne j$ 
        (interference)   in the process  $e^+e^-\to W^+W^-$. Here, $V\in\{\gamma,Z\}$.
        The ``\checkmark" ({\it check mark}) represents the presence  and ``---" ({\it big-dash})
        corresponds to  absence.}
    \renewcommand{\arraystretch}{1.550}
    \begin{tabular*}{\textwidth}{@{\extracolsep{\fill}}cccccccccccc@{}}\hline
        Parameters & $\sigma$ & $\sigma\times A_x$ & $\sigma\times A_y$ & $\sigma\times A_z$ &$\sigma\times A_{xy}$  
        &$\sigma\times A_{xz}$  & $\sigma\times A_{yz}$ & $\sigma\times A_{x^2-y^2}$ & $\sigma\times A_{zz}$ & $\sigma\times A_{fb}$ \\\hline
        $\Delta g_1^V$ & \checkmark & \checkmark & --- & \checkmark & --- & \checkmark & --- & \checkmark & \checkmark & \checkmark \\
        $g_4^V $& --- & --- & \checkmark & --- & \checkmark & --- & \checkmark & --- & --- & --- \\
        $g_5^V $& \checkmark & \checkmark & --- & \checkmark & --- & \checkmark & --- & \checkmark & \checkmark & \checkmark \\
        $\lambda^V $& \checkmark & \checkmark & --- & \checkmark & --- & \checkmark & --- & \checkmark & \checkmark & \checkmark \\
        $\wtil{\lambda^V}$ & --- & --- & \checkmark & --- & \checkmark & --- & \checkmark & --- & --- & --- \\
        $\Delta\kappa^V$ & \checkmark & \checkmark & --- & \checkmark & --- & \checkmark & --- & \checkmark & \checkmark & \checkmark \\
        $\wtil{\kappa^V}$ & --- & --- & \checkmark & --- & \checkmark & --- & \checkmark & --- & --- & --- \\
        $(\Delta g_1^V)^2 $& \checkmark & \checkmark & --- & --- & --- & --- & --- & \checkmark & \checkmark & --- \\
        $(g_4^V)^2$& \checkmark & --- & --- & --- & --- & --- & --- & \checkmark & \checkmark & --- \\
        $(g_5^V)^2 $& \checkmark & --- & --- & --- & --- & --- & --- & \checkmark & \checkmark & --- \\
        $(\lambda^V)^2$& \checkmark & \checkmark & --- & --- & --- & --- & --- & \checkmark & \checkmark & --- \\
        $(\wtil{\lambda^V})^2$& \checkmark & \checkmark & --- & --- & --- & --- & --- & \checkmark & \checkmark & --- \\
        $(\Delta\kappa^V)^2 $& \checkmark & \checkmark & --- & --- & --- & --- & --- & \checkmark & \checkmark & --- \\
        $(\wtil{\kappa^V})^2 $& \checkmark & \checkmark & --- & --- & --- & --- & --- & \checkmark & \checkmark & --- \\
        $\Delta g_1^V g_4^V $& --- & --- & --- & --- & --- & --- & \checkmark & --- & --- & --- \\
        $\Delta g_1^V g_5^V $& --- & --- & --- & \checkmark & --- & --- & --- & --- & --- & \checkmark \\
        $\Delta g_1^V \lambda^V $& \checkmark & \checkmark & --- & --- & --- & --- & --- & \checkmark & \checkmark & --- \\
        $\Delta g_1^V \wtil{\lambda^V} $& --- & --- & \checkmark & --- & \checkmark & --- & --- & --- & --- & --- \\
        $\Delta g_1^V \Delta\kappa^V $& \checkmark & \checkmark & --- & --- & --- & --- & --- & \checkmark & \checkmark & --- \\
        $\Delta g_1^V \wtil{\kappa^V} $& --- & --- & \checkmark & --- & \checkmark & --- & --- & --- & --- & --- \\
        $g_4^V g_5^V $& --- & --- & --- & --- & \checkmark & --- & --- & --- & --- & --- \\
        $g_4^V \lambda^V $& --- & --- & --- & --- & --- & --- & \checkmark & --- & --- & --- \\
        $g_4^V \wtil{\lambda^V} $& --- & --- & --- & \checkmark & --- & \checkmark & --- & --- & --- & \checkmark \\
        $g_4^V \Delta\kappa^V $& --- & --- & --- & --- & --- & --- & \checkmark & --- & --- & --- \\
        $g_4^V \wtil{\kappa^V} $& --- & --- & --- & \checkmark & --- & \checkmark & --- & --- & --- & \checkmark \\
        $g_5^V \lambda^V $& --- & --- & --- & \checkmark & --- & \checkmark & --- & --- & --- & \checkmark \\
        $g_5^V \wtil{\lambda^V} $& --- & --- & --- & --- & --- & --- & \checkmark & --- & --- & --- \\
        $g_5^V \Delta\kappa^V $& --- & --- & --- & \checkmark & --- & \checkmark & --- & --- & --- & \checkmark \\
        $g_5^V \wtil{\kappa^V} $& --- & --- & --- & --- & --- & --- & \checkmark & --- & --- & --- \\
        $\lambda^V \wtil{\lambda^V}$ & --- & --- & \checkmark & --- & \checkmark & --- & --- & --- & --- & --- \\
        $\lambda^V \Delta\kappa^V $& \checkmark & \checkmark & --- & --- & --- & --- & --- & \checkmark & \checkmark & --- \\
        $\lambda^V\wtil{\kappa^V}  $& --- & --- & \checkmark & --- & \checkmark & --- & --- & --- & --- & --- \\
        $\wtil{\lambda^V} \Delta\kappa^V $& --- & --- & \checkmark & --- & \checkmark & --- & --- & --- & --- & --- \\
        $\wtil{\lambda^V}\wtil{\kappa^V}  $& \checkmark & \checkmark & --- & --- & --- & --- & --- & \checkmark & \checkmark & --- \\
        $\Delta\kappa^V \wtil{\kappa^V} $& --- & --- & \checkmark & --- & \checkmark & --- & --- & --- & --- & --- \\
        \hline
    \end{tabular*}
\end{table*}
\begin{table*}
	\caption{\label{tab:Limits-review-OpLag}The list of  posterior   $95~\%$ BCI
			of anomalous couplings $c_i^{\cal O}$ (TeV$^{-2}$)  of effective operators  in Eq.~(\ref{eq:opertaors-dim6}) and their translated limits
			on the couplings $c_i^{{\cal L}_g}$ ($10^{-2}$)
			for $\sqrt{s}=500$ GeV and set of luminosities and beam polarizations (ILC projected) in
			{\tt Binned} case   from MCMC with the same notation used in Table~\ref{tab:Limits-Lag}.}
	\renewcommand{\arraystretch}{1.50}
	\begin{tabular*}{\textwidth}{@{\extracolsep{\fill}}ccccccc@{}}\hline
		$(\eta_3,\xi_3)$& $(0.0,0.0)$ & \multicolumn{2}{c}{$(\pm 0.8,0)$} & \multicolumn{2}{c}{$(\pm 0.8,\mp 0.3)$} & $(\pm 0.8,\mp 0.6)$ \\ \hline
		${\cal L}$ & $4$ ab$^{-1}$ & $100$ fb$^{-1}$ & $4$ ab$^{-1}$  & $100$ fb$^{-1}$ & $3.2$ ab$^{-1}$ & $3.2$ ab$^{-1}$ \\ \hline
		
		$\frac{c_{WWW}}{\Lambda^2}             $&$ _{-0.52  }^{+0.44  }$&$ _{-1.1   }^{+1.2    }$&$ _{-0.41 }^{+0.41  }$&$ _{-1.0   }^{+1.1    }$&$ _{-0.41    }^{+0.42     }$&$ _{-0.41     }^{+0.42  }$\\  \hline
		$\frac{c_{W}}{\Lambda^2}               $&$ _{-0.50  }^{+0.59  }$&$ _{-0.85  }^{+0.70   }$&$ _{-0.29 }^{+0.27  }$&$ _{-0.75  }^{+0.59   }$&$ _{-0.28    }^{+0.26     }$&$ _{-0.27     }^{+0.25  }$\\  \hline
		$\frac{c_{B}}{\Lambda^2}               $&$ _{-1.1   }^{+0.81  }$&$ _{-1.1   }^{+0.81   }$&$ _{-0.25 }^{+0.23  }$&$ _{-0.77  }^{+0.64   }$&$ _{-0.20    }^{+0.19     }$&$ _{-0.17     }^{+0.16  }$\\  \hline
		$\frac{c_{\widetilde{WWW}}}{\Lambda^2} $&$ _{-0.35  }^{+0.35  }$&$ _{-1.0   }^{+1.1    }$&$ _{-0.37 }^{+0.37  }$&$ _{-0.97  }^{+0.96   }$&$ _{-0.38    }^{+0.38     }$&$ _{-0.37     }^{+0.38  }$\\  \hline
		$\frac{c_{\widetilde{W}}}{\Lambda^2}   $&$ _{-3.8   }^{+3.9   }$&$ _{-5.7   }^{+5.9    }$&$ _{-1.3  }^{+1.3   }$&$ _{-4.2   }^{+4.3    }$&$ _{-0.97    }^{+0.97     }$&$ _{-0.69     }^{+0.69  }$\\  \hline\hline
		$\lambda^{\gamma}            		$&$ _{-0.21  }^{+0.18  }$&$ _{-0.47  }^{+0.51   }$&$ _{-0.17 }^{+0.17  }$&$ _{-0.42  }^{+0.47   }$&$ _{-0.17    }^{+0.18     }$&$ _{-0.17     }^{+0.17  }$\\  \hline 
		$\widetilde{\lambda^{\gamma}}		$&$ _{-0.15  }^{+0.14  }$&$ _{-0.43  }^{+0.43   }$&$ _{-0.15 }^{+0.15  }$&$ _{-0.40  }^{+0.40   }$&$ _{-0.16    }^{+0.16     }$&$ _{-0.16     }^{+0.16  }$\\  \hline
		$\Delta\kappa^{\gamma}       		$&$ _{-0.21  }^{+0.15  }$&$ _{-0.38  }^{+0.27   }$&$ _{-0.11 }^{+0.11  }$&$ _{-0.33  }^{+0.24   }$&$ _{-0.12    }^{+0.11     }$&$ _{-0.11     }^{+0.10  }$\\  \hline
		$\widetilde{\kappa^{\gamma}} 		$&$ _{-1.2   }^{+1.3   }$&$ _{-1.9   }^{+1.9    }$&$ _{-0.42 }^{+0.42  }$&$ _{-1.4   }^{+1.4    }$&$ _{-0.31    }^{+0.31     }$&$ _{-0.22     }^{+0.22  }$\\  \hline
		$\Delta g_1^Z                		$&$ _{-0.21  }^{+0.25  }$&$ _{-0.35  }^{+0.29   }$&$ _{-0.12 }^{+0.11  }$&$ _{-0.31  }^{+0.25   }$&$ _{-0.12    }^{+0.11     }$&$ _{-0.11     }^{+0.10  }$\\  \hline
		$\Delta\kappa^Z              		$&$ _{-0.23  }^{+0.29  }$&$ _{-0.30  }^{+0.28   }$&$ _{-0.10 }^{+0.10  }$&$ _{-0.25  }^{+0.22   }$&$ _{-0.092   }^{+0.087    }$&$ _{-0.085    }^{+0.080 }$\\  \hline
		$\widetilde{\kappa^Z}        		$&$ _{-0.36  }^{+0.35  }$&$ _{-0.55  }^{+0.54   }$&$ _{-0.12 }^{+0.12  }$&$ _{-0.39  }^{+0.39   }$&$ _{-0.089   }^{+0.090    }$&$ _{-0.064    }^{+0.064 }$\\  \hline
	\end{tabular*}
\end{table*}
\section{Note on linear approximation}\label{appendix:b}
    If the cross section $\sigma$ is express as a function of couplings $c_i$ as, 
    \begin{equation}
    \sigma=\sigma_0 + \sum_i\sigma_i\times c_i + \sum_{i,j}\sigma_{ij}\times c_i c_j,
    \end{equation}
    linear approximation for the BSM operator will be possible if the quadratic contributions are much
    smaller than the linear contribution, i.e.,
    \begin{equation}
    |\sigma_i\times c_i| \gg |\sigma_{ii}\times c_i^2|,~~~\text{or}~~|c_i|\ll \frac{\sigma_i}{\sigma_{ii}}.
    \end{equation}
    As an example, consider the $\lambda^Z$ dependent unpolarized  cross section given by
    \begin{eqnarray}\label{eq:lalz_unpol}
    \sigma(0.0,0.0)&=& 1037.  + 57. \times \lambda^Z + 12241. \times(\lambda^Z)^2.
    \end{eqnarray}
    The linear approximation is valid for $|\lambda^Z|\ll 0.004 $. However, the limit on $\lambda^Z$   is $\pm 0.36$ at $1\sigma$ level at $100$ 
    fb$^{-1}$ ($2\%$ systematic is used) assuming linear approximation of Eq.~(\ref{eq:lalz_unpol}), which is  much beyond the validity of the linear 
    approximation.  To derive a sensible limit one needs to include the quadratic term which appears 
    at ${\cal O}(\Lambda^{-4})$. However, at   ${\cal O}(\Lambda^{-4})$ one also has the contribution from dimension-$8$ operators
    at linear order. Our present analysis includes quadratic contributions in dimension-$6$ operators and does not 
    include dimension-$8$ contributions to compare our result with the current LHC constrain, Table~\ref{tab:aTGC_constrain_form_collider}.
    However, at higher luminosity ($4$ ab$^{-1}$)  we obtain limits on $\lambda^Z$ to be $10^{-3}$ using binned observables; see Table~\ref{tab:Limits-review-OpLag}. In this range of couplings, the linear terms dominate over the quadratic terms and, hence, linear approximation becomes valid. At high luminosity, thus, our analysis effectively considers only ${\cal O}(\Lambda^{-2})$ terms in the observables. 

\section{Combining  beam polarization with its opposite values}\label{appendix:c}
\begin{figure*}
	\centering
    \includegraphics[width=0.505\textwidth]{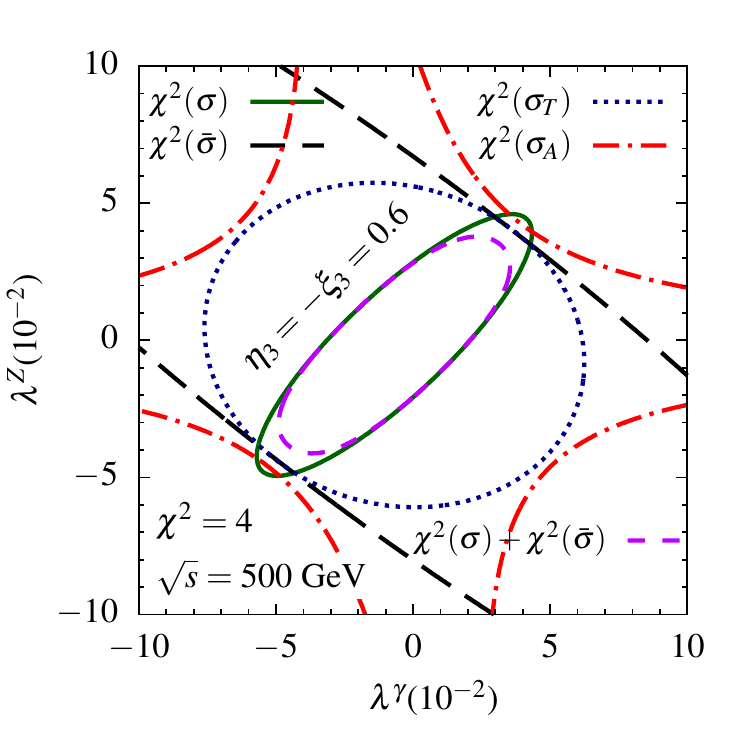}
    \includegraphics[width=0.488\textwidth]{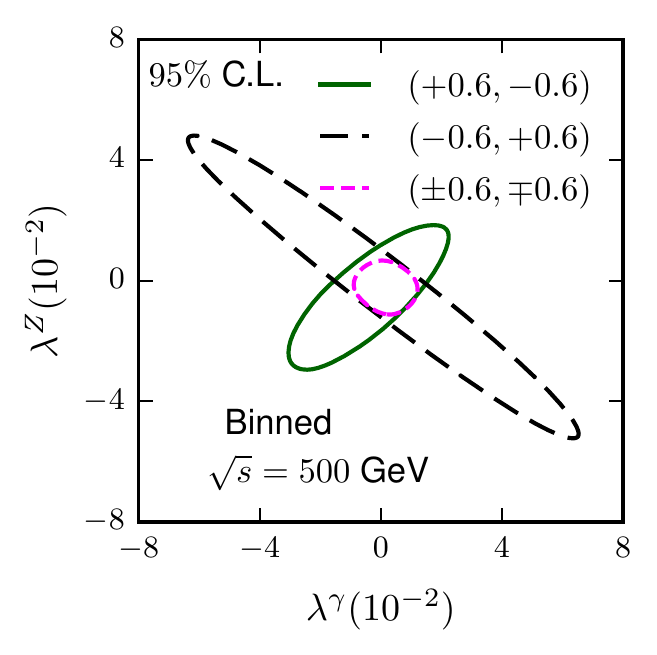}
    \caption{\label{fig:beampol-combine}  The $\chi^2=4$ contours of the unbinned cross section   $\sigma=\sigma(+\eta_3,+\xi_3)$ in {\it solid}/green lines,
 $\bar{\sigma}=\sigma(-\eta_3,-\xi_3)$ in {\it big-dashed}/black lines, $\sigma_T=\sigma(+\eta_3,+\xi_3)+\sigma(-\eta_3,-\xi_3)$ in {\it dotted}/blue line,  $\sigma_A=\sigma(+\eta_3,+\xi_3)-\sigma(-\eta_3,-\xi_3)$ in {\it dash-dotted}/red line and the combined $\chi^2$ of $\sigma$ and
 $\bar{\sigma}$ in {\it dashed}/magenta lines    for polarization $(\eta_3,\xi_3)=(+0.6,-0.6)$ on  $\lambda^\gamma$--$\lambda^Z$ plane are shown in the {\em left-panel}. The $95~\%$ C.L. contours from simultaneous
 analysis in $\lambda^\gamma$--$\lambda^Z$ plane for the beam polarization $(+0.6,-0.6)$, $(-0.6,+0.6)$ and their combined one $(\pm0.6,\mp0.6)$
 are shown in the {\em right-panel} using all the binned observables, i.e., in the {\tt Binned} case. The analyses are done for  $\sqrt{s}=500$ GeV and ${\cal L}=50$ fb$^{-1}$ luminosity to each beam polarization set.} 
\end{figure*}
To reduce the systematic errors in analysis due to luminosity, the beam polarizations 
are flipped between  two opposite choices frequently giving half the total luminosity to both
the polarization choices in an $e^+$--$e^-$ collider.
One can, in principle, use the observables, e.g., the total cross section ($\sigma_T$) or their difference ($\sigma_A$)
as in Eqs.~(\ref{eq:sigma_T}) \& (\ref{eq:sigma_A}), respectively, or
for the two opposite polarization choices ($\sigma$ \& $\bar{\sigma}$) separately for a suitable analysis. In this work, we have combined  the
opposite beam polarization at the level of $\chi^2$ as given in Eq.~(\ref{eq:beampol-combine})
 not at the level of observables as the former constrains the couplings better than any combinations and of-course the individuals.
 To depict this, we present the $\chi^2=4$ contours 
  of the unbinned cross sections in Fig.~\ref{fig:beampol-combine} ({\em left-panel}) for beam polarization $(+0.6,-0.6)$ ($\sigma$) and $(-0.6,+0.6)$ ($\bar{\sigma}$) and the
combinations $\sigma_T$ and $\sigma_A$   along with the combined $\chi^2$
 in the $\lambda^\gamma$--$\lambda^Z$ plane     for  ${\cal L}=50$ 
fb$^{-1}$ luminosity to each polarization choice as representative.
 A systematic error of $2\%$ is used as a benchmark in the cross section.  
 The nature of the contours can be explained as follows: 
In the $WW$ production, the aTGC contributions appear only in the 
$s$-channel (see Fig.~\ref{fig:Feynman-ee-ww}), where initial state
$e^+e^-$ couples through the  $\gamma/Z$ boson and both left and right chiral electrons contribute
almost equally. The $t$-channel diagram, however, is pure background and  
receives contribution only from left chiral 
electrons. As a result, the  $\bar{\sigma}$ ({\it big-dashed}/black) 
contains more background than $\sigma$ ({\it solid}/green)  leading to a weaker limit on the couplings.
Further,  inclusion of $\bar{\sigma}$ into 
$\sigma_T$ ({\it dotted}/blue) and  $\sigma_A$ ({\it dashed-dotted}/red) reduces the signal to the  background ratio, 
and hence they are less  sensitive to the couplings. The total $\chi^2$  for the combined
beam polarizations  shown in {\it dashed} (magenta) is, of course, the best to constrain the couplings.
This behaviour is reverified with the simultaneous analysis using the binned cross section and 
polarization asymmetries ($72$ observables in the {\tt Binned} case) 
and shown in Fig.~\ref{fig:beampol-combine} ({\em right-panel}) in the same $\lambda^\gamma$--$\lambda^Z$ 
plane showing the  $95~\%$ C.L. contours for beam polarizations 
$(+0.6,-0.6)$, $(-0.6,+0.6)$, and their combinations $(\pm0.6,\mp0.6)$.
Thus, we choose to combine the opposite beam polarization choices at the level 
of $\chi^2$ rather than combining them at the level of observables.

\clearpage
\bibliography{eeww}
\bibliographystyle{utphys}

\end{document}